\documentclass[useAMS]{mn2e}

\usepackage{latexsym,graphicx}%,natbib}

\usepackage{rotating}
\usepackage{color}
\newcommand\kms{{\rm\,km\,s^{-1}}}
\newcommand\msun{\rm\,M_\odot}
\newcommand\lsun{\rm\,L_\odot}
\newcommand\rsun{\rm\,R_\odot}
\newcommand\myr{\msun \, {\rm yr}^{-1}}
\newcommand\hii{H\,{\sc ii} \,}
\newcommand{\MC}{\multicolumn}

\DeclareRobustCommand{\ion}[2]{%
\relax\ifmmode
\ifx\testbx\f
{\mathrm{#1\,\textsc{#2}}}\else {\mathrm{#1\,\mathsc{#2}}}\fi
\else\textup{#1\,{\mdseries\textsc{#2}}}%
\fi}

\def\apgt{\ {\raise-.5ex\hbox{$\buildrel>\over\sim$}}\ }
\def\aplt{\ {\raise-.5ex\hbox{$\buildrel<\over\sim$}}\ }

\title[Discovery of a new Wolf--Rayet star in the LMC]{Discovery of a new Wolf--Rayet star and a candidate star cluster in the Large Magellanic Cloud with {\it Spitzer}}
\author[V.V.Gvaramadze et al.]
       {V. V.~Gvaramadze,$^{1,2}$\thanks{E-mail: vgvaram@mx.iki.rssi.ru} A.-N.~Chen\'{e},$^{3,4,5}$ A. Y.~Kniazev,$^{6,7,1}$ O.~Schnurr,$^{8}$
       \newauthor T.~Shenar,$^{9}$ A.~Sander,$^{9}$ R.~Hainich,$^{9}$ N.~Langer,$^{10}$ W.-R.~Hamann,$^{9}$ Y.-H.~Chu$^{11}$
       \newauthor and R.A.~Gruendl$^{11}$ \\
        $^{1}$Sternberg Astronomical Institute, Lomonosov Moscow State University, Universitetskij Pr. 13, Moscow 119992, Russia\\
        $^{2}$Isaac Newton Institute of Chile, Moscow Branch, Universitetskij Pr. 13, Moscow 119992, Russia \\
        $^{3}$Departamento de F\'{i}sica y Astronom\'{i}a, Universidad de Valpara\'{i}so, Av. Gran Breta\~{n}a 1111,
        Playa Ancha, Casilla 5030, Chile \\
        $^{4}$Departamento de Astronom\'{i}a, Universidad de Concepci\'{o}n, Casilla 160-C, Chile \\
        $^{5}$Gemini Observatory, Northern Operations Center, 670 North A'ohoku Place, Hilo, HI 96720, USA \\
        $^{6}$South African Astronomical Observatory, PO Box 9, 7935 Observatory, Cape Town, South Africa \\
        $^{7}$Southern African Large Telescope Foundation, PO Box 9, 7935 Observatory, Cape Town, South Africa \\
        $^{8}$Leibniz-Institut f\"ur Astrophysik Potsdam (AIP), An der Sternwarte 16, 14482 Potsdam, Germany \\
        $^{9}$Institute for Physics and Astronomy, University Potsdam, 14476 Potsdam, Germany \\
        $^{10}$Argelander-Institut f\"ur Astronomie der Universit\"at Bonn, Auf dem H\"ugel 71, 53121 Bonn, Germany \\
        $^{11}$Astronomy Department, University of Illinois, 1002 W. Green Street, Urbana, IL 61801, USA \\
                       }
\begin{document}

\date{Accepted 2014 May 4. Received 2014 May 4; in original form 2014 January 24}

%\pagerange{\pageref{firstpage}--\pageref{lastpage}} \pubeyar{2002}

\maketitle

\label{firstpage}

\begin{abstract}
We report the first-ever discovery of a Wolf--Rayet (WR) star in
the Large Magellanic Cloud via detection of a circular shell with
the {\it Spitzer Space Telescope}. Follow-up observations with
Gemini-South resolved the central star of the shell into two
components separated from each other by $\approx$2 arcsec (or
$\approx$0.5 pc in projection). One of these components turns out
to be a WN3 star with H and He lines both in emission and
absorption (we named it BAT99\,3a using the numbering system based
on extending the Breysacher et al. catalogue). Spectroscopy of the
second component showed that it is a B0\,V star. Subsequent
spectroscopic observations of BAT99\,3a with the du Pont 2.5-m
telescope and the Southern African Large Telescope revealed that
it is a close, eccentric binary system, and that the absorption
lines are associated with an O companion star. We analyzed the
spectrum of the binary system using the non-LTE Potsdam
Wolf--Rayet (PoWR) code, confirming that the WR component is a
very hot ($\approx$90 kK) WN star. For this star, we derived a
luminosity of $\log L/\lsun =5.45$ and a mass-loss rate of
$10^{-5.8} \, \myr$, and found that the stellar wind composition
is dominated by helium with 20 per cent of hydrogen. Spectroscopy
of the shell revealed an He\,{\sc iii} region centred on BAT99\,3a
and having the same angular radius ($\approx$15 arcsec) as the
shell. We thereby add a new example to a rare class of
high-excitation nebulae photoionized by WR stars. Analysis of the
nebular spectrum showed that the shell is composed of unprocessed
material, implying that the shell was swept-up from the local
interstellar medium. We discuss the physical relationship between
the newly identified massive stars and their possible membership
of a previously unrecognized star cluster.
\end{abstract}


\section{Introduction}
\label{sec:intro}

Massive stars are sources of copious stellar winds of variable
mass-loss rate and velocity, which create circumstellar and
interstellar shells of a wide range of morphologies (Johnson \&
Hogg 1965; Chu 1981; Lozinskaya \& Lomovskij 1982; Heckathorn,
Bruhweiler \& Gull 1982; Chu, Treffers \& Kwitter 1983; Dopita et
al. 1994; Marston 1995; Nota et al. 1995; Gruendl et al. 2000;
Weis 2001; Smith 2007; Gvaramadze, Kniazev \& Fabrika 2010a).
Detection of such shells by means of optical, radio or infrared
(IR) observations provides a useful tool for revealing evolved
massive stars. The mid-IR imaging with the {\it Spitzer Space
Telescope} (Werner et al. 2004) and the {\it Wide-field Infrared
Survey Explorer} ({\it WISE}; Wright et al. 2010) turns out to be
the most effective, resulting in discoveries of many dozens of
luminous blue variable (LBV), Wolf--Rayet (WR) and other massive
transient stars (Gvaramadze et al. 2009, 2010b,c, 2012a, 2014;
Wachter et al. 2010, 2011; Mauerhan et al. 2010; Stringfellow et
al. 2012a,b; Burgemeister et al. 2013). The high angular
resolution of {\it Spitzer} images ($\approx$6 arcsec at
$24\,\mu$m) even allowed us to discover parsec-scale circumstellar
nebulae around several already known evolved massive stars in the
Magellanic Clouds (Gvaramadze, Kroupa \& Pflamm-Altenburg 2010d;
Gvaramadze, Pflamm-Altenburg \& Kroupa 2011a).

The vast majority of nebulae associated with massive stars were
detected far from (known) star clusters where the massive stars
are believed to form (Lada \& Lada 2003). This suggests that
central stars of these nebulae are runaways. Indeed, studies of
massive stars in the field showed that most of them can be traced
back to their parent clusters (e.g. Schilbach \& R\"{o}ser 2008)
or are spatially located not far from known star clusters and
therefore could escape from them (de Wit et al. 2004, 2005;
Gvaramadze \& Bomans 2008; Gvaramadze et al. 2011c, 2013a) because
of dynamical encounters with other massive stars (Poveda, Ruiz \&
Allen 1967; Gies \& Bolton 1986) or binary supernova explosions
(Blaauw 1961; Stone 1991; Eldridge, Langer \& Tout 2011).

There are, however, several instances of circumstellar nebulae
containing (at least in projection) several massive stars within
their confines (Figer et al. 1999; Mauerhan et al. 2010; Wachter
et al. 2010; Gvaramadze \& Menten 2012). Some of them, like the
bipolar nebula around the candidate LBV star MWC\,349A (Gvaramadze
\& Menten 2012) could be produced by evolved massive stars in
runaway multiple systems, while others, like the pair of shells
associated with two WN9h stars WR\,120bb and WR\,120bc
(Burgemeister et al. 2013), are simply projected against the
parent cluster of their central stars (Mauerhan et al. 2010). In
the latter case, the well-defined circular shape of nebulae around
WR\,120bb and WR\,120bc implies that these two WR stars are
located outside the cluster, so that their nebulae are not
affected by stellar winds of other massive members of the cluster.

In this paper, we report the discovery of a new circular shell in
the Large Magellanic Cloud (LMC) with {\it Spitzer} and the
results of follow-up spectroscopic observations of its central
star with Gemini-South. We resolved the star into two components,
one of which turns out to be a WN3 star with absorption
lines\footnote{Note that Neugent, Massey \& Morrell (2012)
erroneously attributed the discovery of this star to Zaritsky et
al. (2004).} (the first-ever extragalactic massive star identified
via detection of a circular shell around it) and the second one a
B0\,V star. (Preliminary results of our study were reported in
Gvaramadze et al. 2012c.) The new IR shell and its central stars
are presented in Section\,\ref{sec:nebula}. Section\,\ref{sec:gem}
describes our spectroscopic follow-up of the central stars and
their spectral classification. Section\,\ref{sec:dup} presents the
results of additional spectroscopic observations of the WN3 star
with the du Pont 2.5-m telescope and the Southern African Large
Telescope (SALT), showing that it is a massive binary system. In
Section\,\ref{sec:mod}, we determine fundamental parameters of the
binary components using the Potsdam Wolf-Rayet (PoWR) model
atmospheres. The origin of the shell is analysed in
Section\,\ref{sec:neb}. The physical relationship between the
newly identified massive stars and their possible membership of a
previously unrecognized star cluster are discussed in
Section\,\ref{sec:rel}.

\section{The new circular shell and its central stars}
\label{sec:nebula}

The new circular shell in the LMC was discovered in archival
imaging data from the {\it Spitzer Space Telescope} Legacy Survey
called ``Surveying the Agents of a Galaxy's Evolution"
(SAGE\footnote{http://sage.stsci.edu/}; Meixner et al. 2006). This
survey provides $\approx$$7\degr$$\times$$7\degr$ images of the
LMC, obtained with the Infrared Array Camera (IRAC, near-IR
wavebands centred at 3.6, 4.5, 5.8 and 8.0\,$\mu$m; Fazio et al.
2004) and the Multiband Imaging Photometer for {\it Spitzer}
(MIPS, mid- and far-IR wavebands centred at 24, 70 and
160\,$\mu$m; Rieke et al. 2004). The resolution of the IRAC images
is $\approx 2$ arcsec and that of the MIPS ones is 6, 18 and 40
arcsec, respectively (Meixner et al. 2006).

\begin{figure*}
\begin{center}
\includegraphics[width=15cm,angle=0]{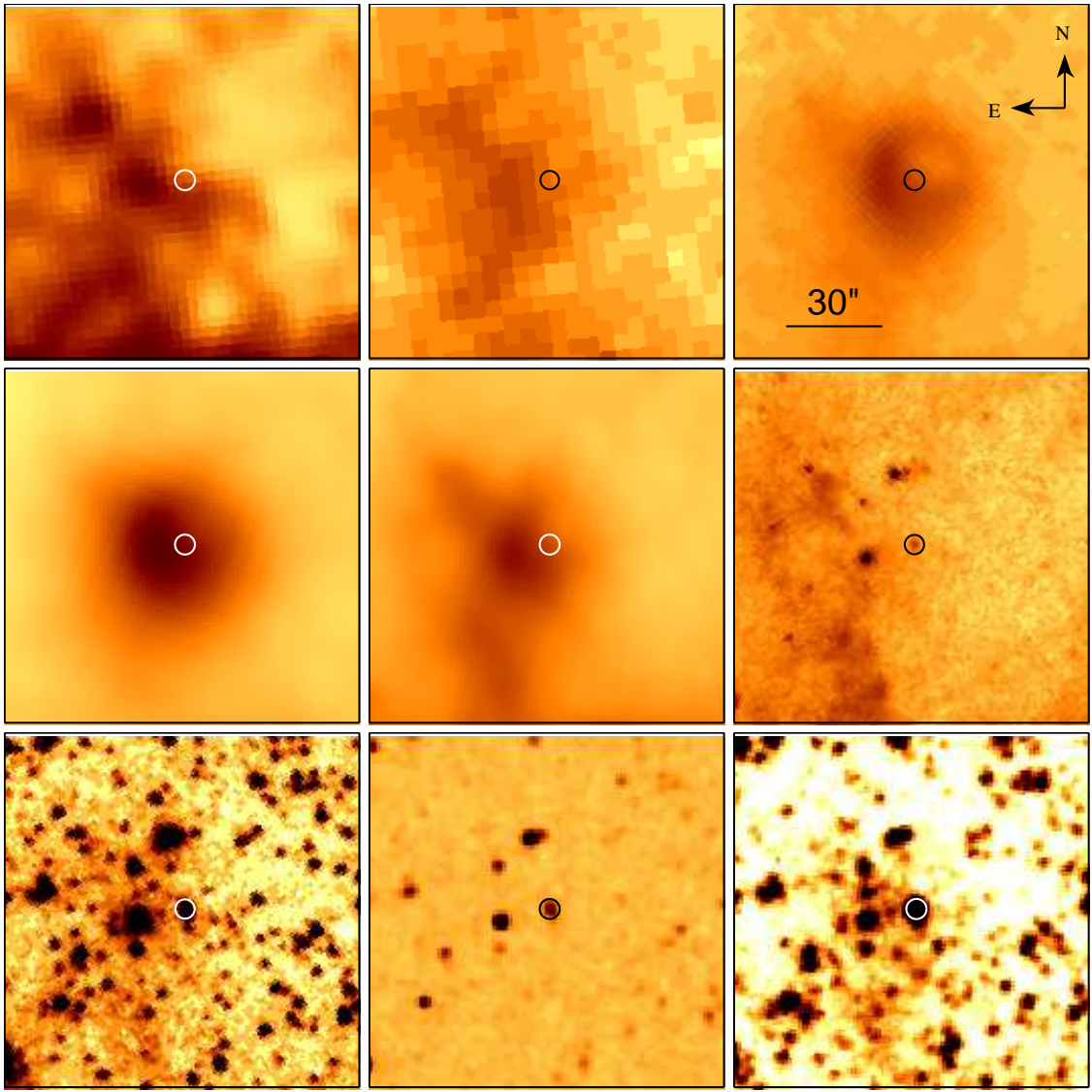}
\end{center}
\caption{From left to right, and from top to bottom: {\it
Herschel} PACS $160\,\mu$m, {\it Spitzer} MIPS 70 and $24\,\mu$m,
{\it WISE} 22 and $12\,\mu$m, {\it Spitzer} IRAC 8 and
3.6\,$\mu$m, 2MASS $J$-band, and DSS-II red-band images of the
region of the LMC containing the new circular shell and its
central star (indicated by a circle). The orientation and the
scale of the images are the same. At the distance of the LMC of 50
kpc, 30 arcsec corresponds to $\approx$7.2 pc.} \label{fig:neb}
\end{figure*}

Fig.\,\ref{fig:neb} shows the {\it Herschel} Space Observatory
(Pilbratt et al. 2010) PACS\footnote{PACS=Photodetector Array
Camera and Spectrometer (Poglitsch et al. 2010).} $160\,\mu$m,
MIPS 70 and $24\,\mu$m, {\it WISE} 22 and $12\,\mu$m, IRAC 8 and
3.6\,$\mu$m, Two-Micron All Sky Survey (2MASS) $J$-band (Skrutskie
et al. 2006), and Digitized Sky Survey II (DSS-II) red-band
(McLean et al. 2000) images of the region containing the circular
shell and its central star (indicated by a circle). Like the
majority of other compact nebulae discovered with {\it Spitzer}
(Gvaramadze et al. 2010a; Mizuno et al. 2010; Wachter et al.
2010), the shell is most prominent at $24 \, \mu$m. It is also
discernible in the {\it WISE} $22 \, \mu$m image and to a lesser
extent at $12 \, \mu$m\footnote{The resolution of these images is
$\approx$12 and 6 arcsec, respectively.}. There is also a gleam of
70\,$\mu$m emission possibly associated with the shell, but the
poor resolution at this wavelength and the fore/background
contamination make this association unclear. In the 24\,$\mu$m
image the shell appears as a ring-like structure with an angular
radius of $\approx$15 arcsec and enhanced brightness along the
eastern rim. At the distance of the LMC of 50 kpc (Gibson 2000), 1
arcsec corresponds to $\approx$0.24 pc, so that the linear radius
of the shell is $\approx$3.6 pc.

The central star is offset by several arcsec from the geometric
centre of the shell, being closer to its brightest portion. This
displacement and the brightness asymmetry could be understood if
the shell impinges on a more dense ambient medium in the eastern
direction. The presence of the dense material on the eastern side
of the shell could be inferred from all but the $J$-band images,
showing a diffuse emission to the east of the star (see also
Section\,\ref{sec:neb}). This region is classified in Bica et al.
(1999) as `NA', i.e. a stellar system clearly related to emission
(named in the SIMBAD data base as BSDL\,161). Alternatively, the
enhanced brightness of the eastern side of the shell and the
displacement of the star might be caused by the stellar motion in
the west-east direction (we discuss both possibilities in
Section\,\ref{sec:rel}).

\begin{figure}
\begin{center}
\includegraphics[width=7.5cm,angle=0]{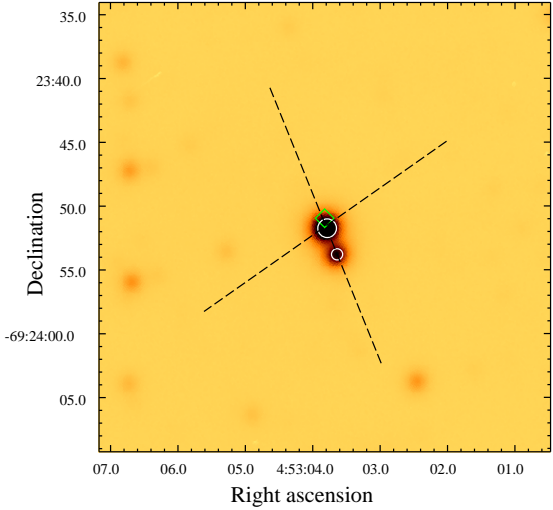}
\end{center}
\caption{GMOS $g'$-band acquisition image of the central star of
the circular shell showing that it is composed of two components,
separated from each other by $\approx$2 arcsec or $\approx0.5$ pc
in projection. The new WR star BAT99\,3a (star\,1) is marked by a
large circle. A small circle marks star\,2 (B0\,V). A diamond
indicates the position of a third star (star\,3), revealed by the
$JHK_{\rm s}$ survey of the Magellanic Clouds by Kato et al.
(2007). The orientations of the spectrograph slit in our two
Gemini observations are shown by dashed lines. See the text for
details.} \label{fig:acq}
\end{figure}

The $g'$-band acquisition image obtained during our spectroscopic
follow-up with the Gemini Multi-Object Spectrograph (GMOS; Hook et
al. 2004) resolved the central star into two components (hereafter
star\,1 and star\,2; marked in Fig.\,\ref{fig:acq} by a large and
small white circle, respectively) separated from each other by
$\approx$2 arcsec or $\approx$0.5 pc in projection. Using the
VizieR catalogue access
tool\footnote{http://webviz.u-strasbg.fr/viz-bin/VizieR}, we
searched for photometry of these stars and found four catalogues
which provide their $UBV$ (Braun 2001), $UBVI$ (Zaritsky et al.
2004), $JHK_{\rm s}$ (Kato et al. 2007), and IRAC band (SAGE LMC
and SMC\footnote{SMC=Small Magellanic Cloud.} IRAC Source
Catalog\footnote{http://irsa.ipac.caltech.edu/applications/Gator/})
magnitudes separately. Interestingly, the $JHK_{\rm s}$ survey of
the Magellanic Clouds by Kato et al. (2007)\footnote{This survey,
carried out with the Infrared Survey Facility 1.4-m telescope at
Sutherland, the South African Astronomical Observatory, resolves
neighboring stars separated by more than $\sim$0.3 pc and its
$10\sigma$ limiting magnitudes are 18.8, 17.8 and 16.6 mag at $J,
H$ and $K_{\rm s}$, respectively, which are fainter by $\approx$2
mag than the two recent near-IR surveys, the 2MASS and the Deep
Near Infrared Survey of the Southern Sky (DENIS; DENIS Consortium,
2005).} resolved star\,1 into two components separated from each
other by 1 arcsec or $\approx$0.24 pc in projection. For the
brightest of these two stars we keep the name star\,1, while the
second one, shown in Fig.\,\ref{fig:acq} by a green diamond, we
call star\,3. The details of the three stars are summarized in
Table\,\ref{tab:phot}. A possible relationship between these stars
is discussed in Section\,\ref{sec:rel}.

\begin{table}
  \caption{Details of three stars in the centre of the circular shell.
  The spectral types, SpT, of stars\,1 (BAT99\,3a) and 2 are based on our spectroscopic
  observations, while that of star\,3 is inferred from the $JHK_{\rm s}$ photometry
  (see the text for details) and therefore should be considered with caution. The
  $UBVI$ photometry is from Zaritsky et al. (2004).   The coordinates and $JHK_{\rm s}$
  photometry are from Kato et al. (2007). The IRAC photometry is from the SAGE LMC and
  SMC IRAC Source Catalog.
  }
  \label{tab:phot}
  \renewcommand{\footnoterule}{}
  %\begin{center}
  \begin{minipage}{\textwidth}
    \begin{tabular}{lccc}
      \hline
       & star\,1 & star\,2 & star\,3 \\
      \hline
      SpT & WN3b+O6\,V & B0\,V & O8.5\,V (?) \\
      $\alpha$ (J2000) & $04^{\rm h} 53^{\rm m} 03\fs76$ & $04^{\rm h} 53^{\rm m} 03\fs60$ & $04^{\rm h} 53^{\rm m} 03\fs86$ \\
      $\delta$ (J2000) & $-69\degr 23\arcmin 51\farcs8$ & $-69\degr 23\arcmin 53\farcs8$ & $-69\degr 23\arcmin 50\farcs9$ \\
      $U$ (mag) & 13.167$\pm$0.041 & 14.502$\pm$0.058 & --- \\
      $B$ (mag) & 14.210$\pm$0.035 & 15.255$\pm$0.044 & --- \\
      $V$ (mag) & 14.391$\pm$0.099 & 15.298$\pm$0.231 & --- \\
      $I$ (mag) & 14.402$\pm$0.102 & 15.441$\pm$0.046 & --- \\
      $J$ (mag) & 14.630$\pm$0.010 & 15.550$\pm$0.020 & 16.420$\pm$0.040 \\
      $H$ (mag) & 14.530$\pm$0.040 & 15.600$\pm$0.020 & 15.920$\pm$0.030 \\
      $K_{\rm s}$ (mag) & 14.610$\pm$0.040 & 15.620$\pm$0.050 & 15.670$\pm$0.070 \\
      $[3.6]$ (mag) & 14.053$\pm$0.052 & 15.311$\pm$0.086 & --- \\
      $[4.5]$ (mag) & 14.120$\pm$0.047 & 15.377$\pm$0.077 & --- \\
      $[5.8]$ (mag) & 13.764$\pm$0.069 & --- & --- \\
      $[8.0]$ (mag) & 13.439$\pm$0.100 & --- & --- \\
      \hline
    \end{tabular}
    \end{minipage}
    %\end{center}
    \end{table}

\section{Spectroscopic follow-up with Gemini-South}
\label{sec:gem}

\subsection{Observations and data reduction}

To determine the spectral type of the central star associated with
the $24\,\mu$m circular shell, we used the Poor Weather time at
Gemini-South under the program-ID GS-2011A-Q-88. The spectroscopic
follow-up was performed with the GMOS in a long-slit mode with a
slit width of 0.75 arcsec. Two spectra were collected on 2011
February 9 and March 5. The log of these and five subsequent
observations with the du Pont telescope and the SALT (see
Section\,\ref{sec:dup}) is listed in Table\,\ref{tab:log}.

The acquisition image obtained during the first observation
resolved the central star into two components, and the decision
was made then by the observer to put only the brightest star
(star\,1; see Table\,\ref{tab:phot}) in the slit (with a position
angle of PA=125$\degr$, measured from north to east; see
Fig.\,\ref{fig:acq}). A quick extraction of the spectrum allowed
us to classify this star as a WN3 star with absorption lines (see
Section\,\ref{sec:WR}), hence we decided to re-observe it to
search for shifts in the radial velocity as a result of possible
binarity. The second spectrum was taken with the slit aligned
along stars\,1 and 2 (PA=22$\degr$).

The first spectrum was obtained under fairly good conditions: with
only some clouds and a $\approx$1 arcsec seeing. The aimed
signal-to-noise ratio (S/N) of $\approx$200 was achieved with a
total exposure time of 450 seconds. For the second observation, in
order to get fairly good S/N on the fainter star (star\,2), we
preferred to double the exposure time, which resulted in the S/N
of 350 in the spectrum of star\,1. Calibration-lamp (CuAr) spectra
and flat-field frames were provided by the Gemini Facility
Calibration Unit (GCal).

\begin{table*}
  \caption{Journal of the observations.}
  \label{tab:log}
  \renewcommand{\footnoterule}{}
  \begin{tabular}{llccccccccc}
      \hline
      Spectrograph & Date & Grating & Spectral scale & Spatial scale    & Slit   & Slit PA & Spectral range \\
      & & (l mm$^{-1}$) & (\AA \, pixel$^{-1}$) & (arcsec pixel$^{-1}$) & arcsec & ($\degr$) & (\AA) \\
      \hline
      GMOS (Gemini)  & 2011 February 9  &  600   & 0.46 & 0.073 & 0.75  & 125 & 3800$-$6750 \\
      GMOS (Gemini)  & 2011 March 5     &  600   & 0.46 & 0.073 & 0.75  & 22  & 3800$-$6750 \\
      B\&C (du Pont) & 2012 December 6  & 1200   & 0.79 & 0.70  & 1.00  & 90  & 3815$-$5440 \\
      B\&C (du Pont) & 2012 December 13 & 1200   & 0.79 & 0.70  & 1.00  & 90  & 3815$-$5440 \\
      B\&C (du Pont) & 2012 December 28 & 1200   & 0.79 & 0.70  & 1.00  & 25  & 3815$-$5440 \\
      RSS (SALT)     & 2013 January  26 & 900    & 0.99 & 0.25  & 1.25$\times$480 & 130 & 3580$-$6700 \\
      RSS (SALT)     & 2013 January  28 & 900    & 0.98 & 0.25  & 1.25$\times$480 & 130 & 4170$-$7300 \\
      \hline
    \end{tabular}
    \end{table*}

\begin{figure*}
\begin{center}
\includegraphics[width=9cm,angle=270,clip=]{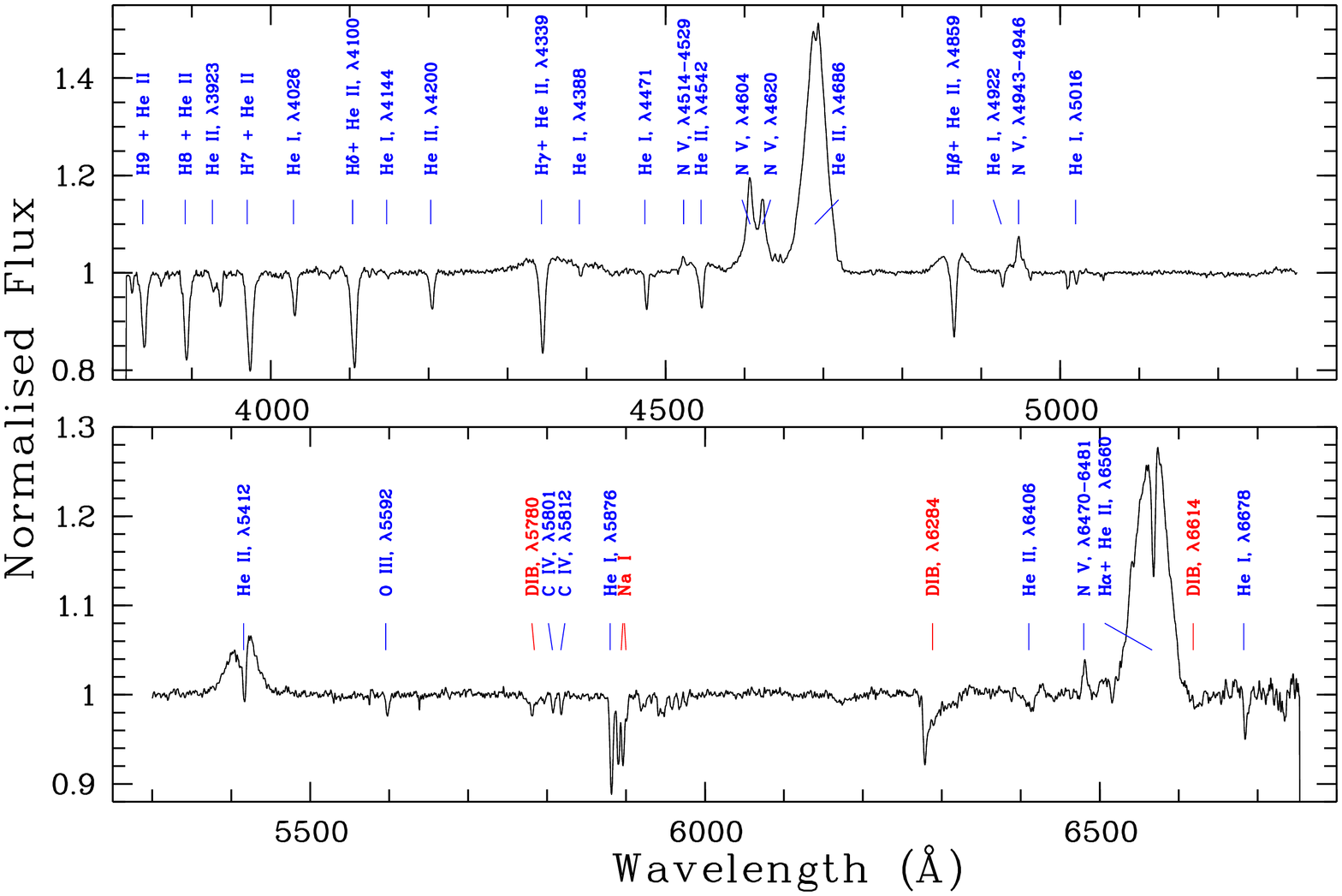}
\end{center}
\caption{Normalized spectrum of the new WR star BAT99\,3a
(star\,1) in the LMC observed with Gemini-South on 2011 March 5
with principal lines and most prominent DIBs indicated.}
\label{fig:spec}
\end{figure*}

\begin{figure*}
\begin{center}
\includegraphics[width=9cm,angle=270,clip=]{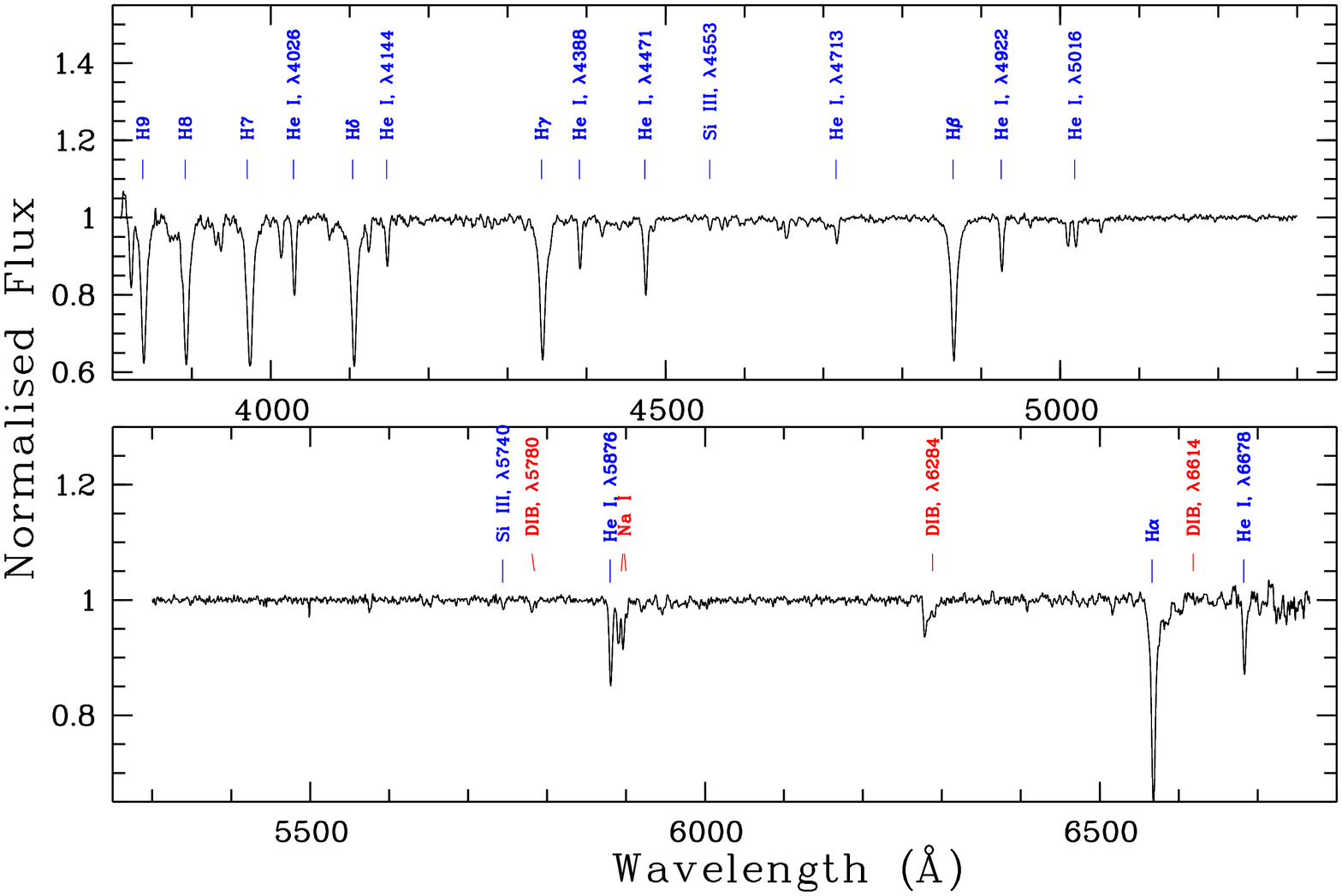}
\end{center}
\caption{Normalized spectrum of star\,2 (B0\,V) observed with
Gemini-South with principal lines and most prominent DIBs
indicated.} \label{fig:spec2}
\end{figure*}

The bias subtraction, flat-fielding, wavelength calibration and
sky subtraction were executed with the {\sc gmos} package in the
{\sc gemini} library of the {\sc iraf}\footnote{{\sc iraf}: the
Image Reduction and Analysis Facility is distributed by the
National Optical Astronomy Observatory (NOAO), which is operated
by the Association of Universities for Research in Astronomy, Inc.
(AURA) under cooperative agreement with the National Science
Foundation (NSF).} software. In order to fill the gaps between
GMOS-S's EEV CCDs, each observation was divided into three
exposures obtained with a different central wavelength, i.e. with
a 5\AA \, shift between each exposure. The extracted spectra were
obtained by averaging the three individual exposures, using a
sigma clipping algorithm to eliminate the effects of cosmic rays.
The B600 grating was used to cover the spectral range
$\lambda\lambda$=3800$-$6750 \AA \, with a reciprocal dispersion
of $\approx$0.5 \AA~pixel$^{-1}$ and the average spectral
resolution FWHM of $\approx$4.08 \AA. The accuracy of the
wavelength calibration estimated by measuring the wavelength of
ten lamp emission lines is 0.065\,\AA. A spectrum of the white
dwarf LTT\,3218 was used for flux calibration and removing the
instrument response. The resulting spectra of stars\,1 and 2 are
shown in Figs\,\ref{fig:spec} and \ref{fig:spec2}, respectively.


Note that star\,1 and the standard star were observed at different
air masses, during different nights, and were not observed with
the longslit aligned with the parallactic angle. These make the
absolute flux calibration of the Gemini spectra very uncertain.

Equivalent widths (EWs), full widths at half-maximum (FWHMs) and
heliocentric radial velocities (RVs) of main lines in the spectra
of stars\,1 and 2 (measured applying the {\sc midas} programs; see
Kniazev et al. 2004 for details) are summarized in
Tables\,\ref{tab:int} and \ref{tab:int2}, respectively. All
wavelengths are in air. For EWs and RVs we give their mean values
derived from both spectra, while for measurements of FWHMs we used
the second spectrum alone owing to its better quality.

\begin{table} \centering{ \caption{EWs, FWHMs and RVs of main lines in the spectrum
of the new WR star BAT99\,3a (star\,1). For EWs and RVs we give
their mean values derived from both Gemini spectra, while for
measurements of FWHMs we used the second spectrum alone owing to
its better quality. } \label{tab:int}
\begin{tabular}{lccc}
\hline \rule{0pt}{10pt} $\lambda_{0}$(\AA) Ion     &EW($\lambda$)$^a$ & FWHM($\lambda$)$^a$ & RV \\
 & (\AA) & (\AA) & ($\kms$) \\ \hline
3889\ He\ {\sc ii}\ +\ H8\        &1.35$\pm$0.13   &  7.37$\pm$0.16  & 311$\pm$40  \\
3968\ He\ {\sc ii}\ +\ H7\        &1.73$\pm$0.17   &  8.32$\pm$0.11  & 271$\pm$11  \\
4026\ He\ {\sc ii}\               &0.56$\pm$0.11   &  5.88$\pm$0.15  & 353$\pm$8   \\
4100\ He\ {\sc ii}\ +\ H$\delta$\ &1.67$\pm$0.16   &  8.09$\pm$0.14  & 285$\pm$6  \\
4144\ He\ {\sc i}\                &0.05$\pm$0.03   &  4.86$\pm$0.35  & 307$\pm$50  \\
4200\ He\ {\sc ii}\               &0.47$\pm$0.11   &  5.82$\pm$0.12  & 317$\pm$8   \\
4339\ He\ {\sc ii}\ +\ H$\gamma$\ &1.49$\pm$0.15   &  7.81$\pm$0.11  & 268$\pm$6  \\
4388\ He\ {\sc i}\                &0.14$\pm$0.09   &  5.08$\pm$0.35  & 361$\pm$35   \\
4471\ He\ {\sc i}\                &0.35$\pm$0.08   &  4.12$\pm$0.11  & 315$\pm$6   \\
4542\ He\ {\sc ii}\               &0.62$\pm$0.12   &  6.81$\pm$0.18  & 266$\pm$8   \\
4604\ N\  {\sc v}\                &--2.50$\pm$0.28 & 13.73$\pm$0.51  & 192$\pm$8  \\
4620\ N\  {\sc v}\                &--1.50$\pm$0.20 & 10.50$\pm$2.00  & 219$\pm$11  \\
4686\ He\ {\sc ii}\               &0.02$\pm$0.03$^*$   &  3.28$\pm$0.21$^*$  & 282$\pm$25   \\
4686\ He\ {\sc ii}\               &--16.37$\pm$0.68$^*$ & 31.10$\pm$0.21$^*$  & 280$\pm$15   \\
4859\ He\ {\sc ii}\ +\ H$\beta$\  &1.14$\pm$0.14$^*$   &  7.09$\pm$0.15$^*$  & 244$\pm$10   \\
4859\ He\ {\sc ii}\ +\ H$\beta$\  &--1.52$\pm$0.52$^*$ & 28.87$\pm$4.09$^*$  & 146$\pm$50   \\
4922\ He\ {\sc i}\                &0.11$\pm$0.06   &  3.43$\pm$0.28  & 306$\pm$20   \\
4944\ N\  {\sc v}\                &--0.50$\pm$0.13 &  8.77$\pm$0.43  & 175$\pm$21  \\
5412\ He\ {\sc ii}\               &0.26$\pm$0.07$^*$   &  5.36$\pm$0.26$^*$  & 261$\pm$10   \\
5412\ He\ {\sc ii}\               &--2.41$\pm$0.83$^*$ & 54.91$\pm$2.10$^*$  & 270$\pm$14  \\
5592\ O\ {\sc iii}\               &  0.11$\pm$0.05 &  4.40$\pm$0.21  & 305$\pm$14   \\
5876\ He\ {\sc i}\                &0.50$\pm$0.07   &  4.64$\pm$0.10  & 313$\pm$21   \\
6560\ He\ {\sc ii}\ +\ H$\alpha$\ &0.56$\pm$0.07$^*$   &  5.08$\pm$0.15$^*$  & 211$\pm$20   \\
6563\ He\ {\sc ii}\ +\ H$\alpha$\ &--12.66$\pm$0.86$^*$ & 47.58$\pm$0.79$^*$ & 177$\pm$15  \\
6678\ He\ {\sc i}\                & 0.17$\pm$0.05  & 4.66$\pm$0.30   & 302$\pm$9  \\
\hline \multicolumn{4}{p{8cm}}{$^a$Note that some of the starred
EWs and FWHMs are significantly affected because of the binary
nature of BAT99\,3a. The negative (positive) EWs correspond to
emission (absorption) lines originating from the WR (O) component
of BAT99\,3a (see the text for details).}
\end{tabular}
 }
\end{table}
\begin{table}
\centering{ \caption{EWs, FWHMs and RVs of main lines in the
spectrum of star\,2 (B0\,V).} \label{tab:int2}
\begin{tabular}{lccc}
\hline \rule{0pt}{10pt} $\lambda_{0}$(\AA) Ion  & EW($\lambda$)  &
FWHM($\lambda$) &     RV         \\
 & (\AA) & (\AA) & ($\kms$) \\ \hline
3889\ He\ {\sc i}\ +\ H8\  &  1.03$\pm$0.01 &  7.84$\pm$0.08 &   298$\pm$1   \\
3968\ He\ {\sc i}\ +\ H7\  &  3.95$\pm$0.08 & 10.14$\pm$0.22 &   461$\pm$6   \\
4026\ He\ {\sc i}\         &  1.00$\pm$0.02 &  4.78$\pm$0.07 &   273$\pm$1   \\
4102\ H$\delta$\           &  3.71$\pm$0.10 & 10.18$\pm$0.30 &   288$\pm$7   \\
4144\ He\ {\sc i}\         &  0.55$\pm$0.02 &  4.52$\pm$0.13 &   283$\pm$1   \\
4340\ H$\gamma$\           &  3.71$\pm$0.09 & 10.76$\pm$0.29 &   290$\pm$6   \\
4388\ He\ {\sc i}\         &  0.52$\pm$0.01 &  4.28$\pm$0.06 &   274$\pm$1   \\
4471\ He\ {\sc i}\         &  1.07$\pm$0.03 &  5.53$\pm$0.15 &   243$\pm$2   \\
4553\ Si\ {\sc iii}\       &  0.12$\pm$0.01 &  3.79$\pm$0.22 &   262$\pm$1   \\
4861\ H$\beta$\            &  3.42$\pm$0.06 &  9.75$\pm$0.20 &   257$\pm$4   \\
4922\ He\ {\sc i}\         &  0.68$\pm$0.01 &  4.62$\pm$0.08 &   258$\pm$1   \\
5740\ Si\ {\sc iii}\       &  0.08$\pm$0.01 &  4.15$\pm$0.32 &   231$\pm$1   \\
5876\ He\ {\sc i}\         &  0.71$\pm$0.02 &  4.82$\pm$0.10 &   263$\pm$1   \\
6563\ H$\alpha$\           &  2.68$\pm$0.09 &  7.70$\pm$0.28 &   250$\pm$4   \\
6678\ He\ {\sc i}\         &  0.18$\pm$0.01 &  4.14$\pm$0.06 &   238$\pm$4   \\
\hline
\end{tabular}
 }
\end{table}

\subsection{Spectral classification} \label{sec:spec}

\subsubsection{Star\,1: a new WR star in the LMC}
\label{sec:WR}

The spectrum of star\,1 (see Fig.\,\ref{fig:spec}) is dominated by
strong emission lines of H$\alpha$, He\,{\sc ii} $\lambda 4686$
and N\,{\sc v} $\lambda\lambda 4604, 4620$. The H$\alpha$ and
He\,{\sc ii} $\lambda 4686$ lines show central absorption
reversals. Several weaker emissions of N\,{\sc v} (at
$\lambda\lambda 4514-4529, 4943-4946$, and 6470$-$6481) are also
detected. There are also several lower-intensity broad emission
features with narrow absorption reversals, of which the most
prominent are H$\beta$+He\,{\sc ii} $\lambda 4859$ and He\,{\sc
ii} $\lambda 5412$. None of the emission lines shows P\,Cygni
profiles. Other H, He\,{\sc i} and He\,{\sc ii} lines are almost
purely in absorption. The O\,{\sc iii} $\lambda 5592$ and C\,{\sc
iv} $\lambda\lambda 5801, 5812$ lines are also purely in
absorption. The low interstellar extinction in the direction
towards the LMC is manifested in several weak diffuse interstellar
bands (DIBs). A prominent absorption visible in the spectrum at
6277 \AA \, is telluric in origin.

The presence of the N\,{\sc v} $\lambda\lambda 4604, 4620$
emission lines and the absence of the N\,{\sc iv} $\lambda$4057
one imply that the emission spectrum belongs to a WR star of the
ionization subclass WN3 (Smith, Shara \& Moffat 1996). The FWHM of
the He\,{\sc ii} $\lambda 4686$ (emission) line of 31.1$\pm$0.21
\AA \, slightly exceeds the empirically determined limit of 30 \AA
\, for broad-lined stars (Smith et al. 1996), which implies that
star\,1 can be classified WN3b. Since it is believed that
broad-lined WR stars are true hydrogen-free WN stars (Smith \&
Maeder 1998; cf. Foellmi, Moffat \& Guerrero 2003), one can assume
that the numerous absorption lines in the spectrum of star\,1 are
due to a second (unresolved) star, either a binary companion or a
separate star located on the same line-of-sight. Our subsequent
spectroscopic observations of star\,1 showed that it is a binary
system (see Section\,\ref{sec:dup}). We name star\,1 BAT99\,3a
using the numbering system based on extending the Breysacher et
al. (1999) catalogue, as suggested by Howarth \& Walborn (2012).

An indirect support to the possibility that the spectrum of
BAT99\,3a is a superposition of two spectra comes from the
position of this star on the plot of EW(5412) versus FWHM(4686)
for WN stars in the LMC (see fig.\,14 of Smith et al. 1996), where
it lies in a region occupied by binary and composite stars.

The presence of He\,{\sc ii} absorption lines in the spectrum of
BAT99\,3a suggests that one of its components is an O-type star.
Taken at face value, the EWs of the He\,{\sc i} $\lambda 4471$ and
He\,{\sc ii} $\lambda 4542$ lines (see Table\,\ref{tab:int}) imply
that this star is of type O6.0$\pm0.5$\footnote{While the
existence of an unresolved companion star generally affects the
observed EWs, a subsequent spectral analysis (see
Section\,\ref{sec:mod}) showed that the contribution of the WR
star to the visual continuum is negligible, implying that the
measured EWs of the classification absorption lines are very close
to the intrinsic ones.} (Morgan, Keenan \& Kellman 1943; Conti \&
Alschuler 1971). The same spectral type also follows from the
classification scheme for O stars in the yellow-green
($\lambda\lambda 4800-5420$ \AA) proposed by Kerton, Ballantyne \&
Martin (1999). Using the following equation (Kerton et al. 1999)
\begin{equation}
{\rm SpT}=(4.82\pm0.03)+(7.92\pm0.56){\rm EW(4922)}
\label{eqn:ker}
\end{equation}
and EW(4922)=0.11$\pm$0.06 \AA, one finds SpT=5.7$\pm$0.5,
%$\pm$0.5,
which agrees with the spectral type based on the traditional
classification criteria. The non-detection of the Si\,{\sc iv}
$\lambda 4089$ line in the spectrum of BAT99\,3a implies that the
O-type companion is on the main sequence (Conti \& Altschuller
1971), i.e. an O6.0$\pm$0.5\,V star (cf. Section\,\ref{sec:mod}).

Detailed comparison of the two Gemini spectra of BAT99\,3a taken
24 days apart did not show evidence of significant change in RVs
typical of close binaries. Although this might imply that the WR
and O stars are unrelated to each other members of the same
(unrecognized) star cluster (cf. Section\,\ref{sec:rel}), our
subsequent spectroscopic observations (see Section\,\ref{sec:dup})
showed that the two stars form a binary system. Moreover, we know
that the Gemini spectra were taken outside of the periastron
passage and hence should not show significant RV variability.

\subsubsection{Star\,2}
\label{sec:B}

The spectrum of star\,2 is dominated by H and He\,{\sc i}
absorption lines (see Fig.\,\ref{fig:spec2}). No He\,{\sc ii}
lines are visible in the spectrum, which implies that star\,2 is
of B type. Using the EW(H$\gamma$)$-$absolute magnitude
calibration by Balona \& Crampton (1974) and the measured
EW(H$\gamma$)=3.71$\pm$0.09 \AA, we estimated the spectral type of
this star as B0\,V. The same spectral type also follows from
equation\,(\ref{eqn:ker}). With EW(4922)=0.68$\pm$0.01 \AA \,, we
found SpT$\approx$10, which corresponds to B0\,V stars (Kerton et
al. 1999).

Using the $B$ and $V$ magnitudes of star\,2 from
Table\,\ref{tab:phot} and intrinsic $(B-V)_{0}$ colour of $-$0.26
mag (typical of B0\,V stars; e.g. Martins \& Plez 2006), and
assuming the standard total-to-selective absorption ratio
$R_V=3.1$ (cf. Howarth 1983), we found the visual extinction,
$A_{V}$, towards this star and its absolute visual magnitude,
$M_{V}$, of $\approx$0.67 and $-$3.87 mag, respectively. The
latter estimate agrees well with $M_V$ of B0\,V stars of $-$3.84
mag, derived by extrapolation from the absolute magnitude
calibration of Galactic O stars of Martins \& Plez (2006).

\section{Additional spectroscopic observations with the du Pont
telescope and the SALT} \label{sec:dup}

\begin{figure*}
\begin{center}
\includegraphics[width=15cm,angle=0,clip=]{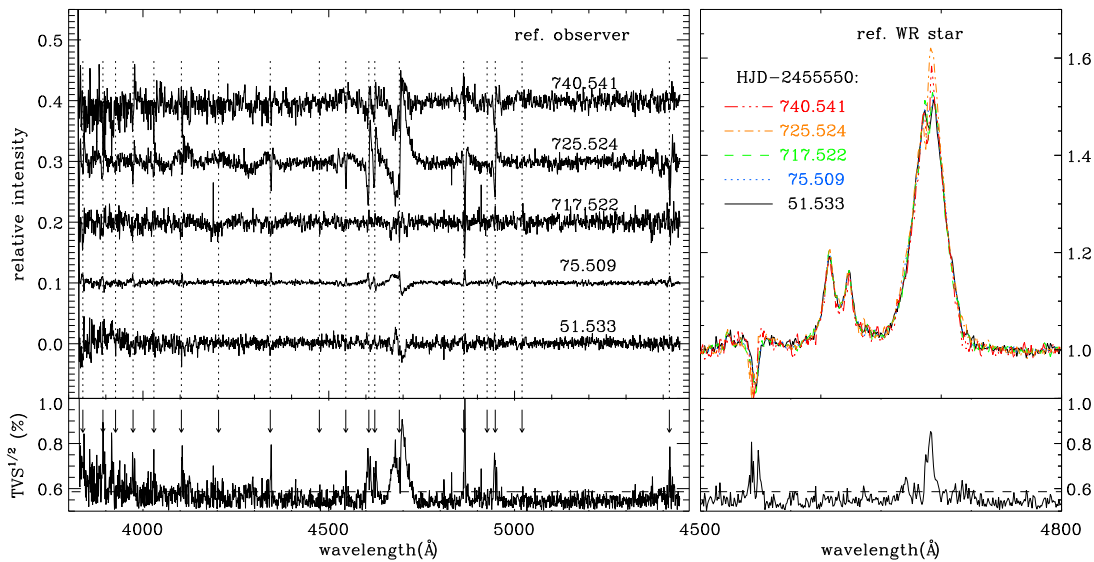}
\end{center}
\caption{Left: The top panel shows a montage of the residuals
(individual spectra minus mean) for our Gemini and du Pont
observations in the observer's reference frame. HJD$-$2455550 is
indicated for each residual, and selected spectral lines are
marked with a dotted line. The bottom panel shows the TVS, which
shows broad double-peaked profiles associated with the emission
lines, and narrow peaks associated with the absorption lines.
Right: The top panel shows the superposition of the five spectra
placed in the reference frame of the WR star. The TVS (bottom
panel) shows only signatures of the motion of the absorption lines
(with respect to the emission ones) and a change in amplitude of
the He\,{\sc ii} lines.} \label{fig:ser}
\end{figure*}

To clarify the possible binary status of BAT99\,3a, five
additional spectra were taken in 2012 December -- 2013 January.

Three spectra were observed as time fillers at the du Pont 2.5-m
telescope (Las Campanas, Chile) on 2012 December 6, 13 and 28,
using the B\&C spectrograph with a slit width of 1 arcsec. A S/N
of $\approx$130 was obtained after a one hour exposure each time.
The wavelength calibration was done using spectra of a HeAr lamp.
The full spectral range is $\lambda\lambda$3815$-$5440 \AA, the
average wavelength resolution is $\approx$0.8 \AA\, pixel$^{-1}$
(FWHM$\approx$2.44 \AA), and the accuracy of the wavelength of ten
lamp emission lines is 0.055 \AA. No spectrophotometric standard
was observed to calibrate the spectra in flux.

Two other spectra were taken with the SALT (Buckley, Swart \&
Meiring 2006; O'Donoghue et al. 2006) on 2013 January 26 and 28,
using the Robert Stobie Spectrograph (RSS; Burgh et al. 2003;
Kobulnicky et al. 2003) in the long-slit mode. The RSS pixel scale
is 0$\farcs$127 and the effective field of view is 8 arcmin in
diameter. We utilized a binning factor of 2, to give a final
spatial sampling of 0$\farcs$254 pixel$^{-1}$. As indicated in
Table~\ref{tab:log}, the volume phase holographic grating GR900
was used mainly to cover the spectral ranges 3580--6700 \AA \, and
4170--7300 \AA \, with a final reciprocal dispersion of $\sim$0.98
\AA\ pixel$^{-1}$ and spectral resolution FWHM of 4--5 \AA.
Observations were done with exposure times of 20 minutes for the
blue spectral range and 10 minutes for the red one. Unfortunately,
they were carried out close to the twilight time and under the
very poor weather conditions. Spectra of Ar comparison arcs were
obtained to calibrate the wavelength scale. Spectrophotometric
standard star LTT\,4364 was observed for the relative flux
calibration. Primary reduction of the data was done with the SALT
science pipeline (Crawford et al. 2010). After that, the bias and
gain corrected and mosaicked long-slit data were reduced in the
way described in Kniazev et al. (2008b). We note that SALT is a
telescope with a variable pupil, so that the illuminating beam
changes continuously during the observations. This makes the
absolute flux/magnitude calibration impossible, even using
spectrophotometric standard stars or photometric standards. At the
same time, the final relative flux distribution is very accurate
and does not depend on the position angle because SALT has an
atmospheric dispersion corrector (O'Donoghue 2002).

Special care was taken for normalizing the continuum of all seven
spectra. The method consists of normalizing all spectra with
respect to a reference spectrum before we fit the continuum on the
overall mean spectrum. The spectrum with the best S/N is always
chosen as the reference. All the other spectra are divided by the
reference, and the results of the divisions are fitted using a
Legendre polynomial of the 4$^{th}$ order. All individual spectra
are divided by the respective fits, so their continuum has the
same shape as the reference spectrum. In our case, we took the
second spectrum from Gemini as the reference. The continuum on the
overall mean spectrum was fitted with a Legendre polynomial of the
8$^{th}$ order and applied to all spectra normalized to the
reference spectrum.

The top left-hand panel of Fig.\,\ref{fig:ser} shows a montage of
the residuals of the Gemini and du Pont spectra subtracted by the
normalized average spectrum. The spectra are organized in
chronological order, and the time (HJD$-$2455550) is indicated for
each spectrum. The du Pont spectra are used to show the line
variations near the periastron passage (see below), and the Gemini
ones are used as a reference when the orbital motion is small. The
residuals show that all absorption and emission lines are varying
with time. The S shape of the residuals at the same wavelength as
the spectral lines is the sign of RV variations. The bottom
left-hand panel shows the square root of the Temporal Variance
Spectrum (TVS; Fullerton, Gies \& Bolton 1996). The spectrum is
significantly variable at the wavelength where the corresponding
TVS$^{1/2}$ value is over the threshold plotted with a dashed
line. We find that all the emission (WR) lines show a double peak
in the TVS, as expected for RV motions.

\begin{table}
  \caption{Mean RVs of N\,{\sc v} $\lambda\lambda$4604,
  4640 emission lines (associated with the WR star) and of the Balmer and
  He\,{\sc ii} absorption lines (associated with the O companion star)
  for seven observations of BAT99 3a.} \label{tab:RVs}
  \renewcommand{\footnoterule}{}
  \begin{center}
   \begin{tabular}{cccc}
      \hline
HJD & RV(N\,{\sc v}) & RV(Balmer) & RV(He\,{\sc ii}) \\
& ($\kms$) & ($\kms$) & ($\kms$) \\
\hline
2455601.533 & 211$\pm$5 & 267$\pm$18 & 267$\pm$66 \\
2455625.509 & 200$\pm$20 & 283$\pm$31 & 298$\pm$49 \\
2456267.522 & 211$\pm$19 & 228$\pm$13 & 273$\pm$21 \\
2456275.524 & 360$\pm$47 & 192$\pm$17 & 220$\pm$18 \\
2456290.541 & 249$\pm$8 & 217$\pm$29 & 241$\pm$24 \\
2456319.300 & 245$\pm$20 & 233$\pm$9 & 265$\pm$59 \\
2456321.301 & 242$\pm$10 & 232$\pm$40 & 233$\pm$55 \\
\hline
    \end{tabular}
    \end{center}
    \end{table}

\begin{figure}
\begin{center}
\includegraphics[width=7.5cm,angle=0,clip=]{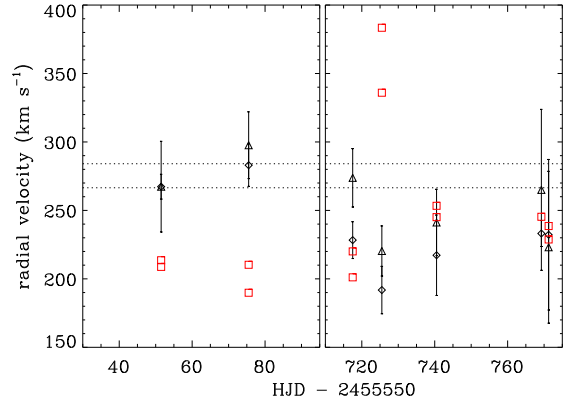}
\end{center}
\caption{RV changes with time in the spectrum of BAT99\,3a. The
average RVs for the Balmer and He\,{\sc ii} absorption lines are
plotted with (black) diamonds and triangles, respectively. The
(red) squares are the RVs for the N\,{\sc v} $\lambda\lambda$4604,
4620 emission lines. The dashed lines are the scatter in RV of the
Balmer lines in star\,2, as observed in the second Gemini
spectrum.} \label{fig:RVc}
\end{figure}

To analyse the RV variability, we used all seven spectra. The RVs
for each line were determined by a non-linear fit of their
profile. Only the He\,{\sc ii} $\lambda\lambda$4686, 4859 emission
lines were not fitted, since they are broad, their profiles are
varying in amplitude and they are blended with absorption lines.
We assume that all the line profiles are close enough to a Voigt
function, even the two N\,{\sc v} emission lines at 4604 and 4620
\AA, for which we get a good fit. The results are listed in
Table\,\ref{tab:RVs} and plotted in Fig.\,\ref{fig:RVc}. We
averaged the RV values for each line series; the absorption Balmer
and He\,{\sc ii} lines are plotted with diamonds and triangles,
respectively, and the (red) squares are the RVs for the two N{\sc
v} emission lines. The dashed lines are the scatter in RV of the
Balmer lines in the spectrum of star\,2, as observed
simultaneously with the second spectrum of BAT99\,3a at Gemini.
The error bars are assumed to be the rms of the individual
measured RVs within a line type (note that our spectral range
includes five Balmer and four He\,{\sc ii} lines well observed).
We prefer to plot the RVs of the N\,{\sc v} lines without the
errors, since only two lines were observed with sufficient S/N.

In the top right-hand panel of Fig.\,\ref{fig:ser}, the observed
spectra are placed into the reference frame of the emission lines
(i.e. the WR star). We see that the He\,{\sc ii} $\lambda$4686
line intensity increases near periastron passage. In the TVS,
shown at the bottom right-hand panel of Fig.\,\ref{fig:ser}, we
see a significant peak centred on 4686 \AA. Its blue side is
dominated by the motion of the O star absorption line, but the red
one shows a variable emission excess, which could be attributed to
the emission of a wind-wind collision zone. Unfortunately, the
absorption lines blended with the He\,{\sc ii} emission lines make
this attribution ambiguous.

The additional spectroscopic observations allow us to confirm that
BAT99\,3a is a binary system, and that the absorption lines
originate in a massive O companion, not in the WR star itself. The
RV variations show that the binary is close and eccentric, and
that we had the chance to observe it near the periastron passage
in December 2012. Our seven spectra, however, are not sufficient
to determine the orbital parameters of the system.

We searched for possible photometric variability of BAT99\,3a
using the Massive Astrophysical Compact Halo Object (MACHO)
project photometric data base (Alcock et al. 1999), but none was
found. The presence of the O-type companion to the WR star implies
that this system could be a source of X-ray emission, arising from
colliding stellar winds (e.g. Usov 1992). Unfortunately, the part
of the LMC containing the circular shell was not observed with
modern X-ray telescopes.

Since the majority of massive stars form close binary systems
(e.g. Chini et al. 2012; Sana et al. 2012, 2013), one can suspect
that star\,2 could be a binary as well. The bad seeing during the
third observation at du Pont, however, did not allow us to
reliably measure RVs for this star and compare them with those
measured in the Gemini spectrum.

\section{BAT99\,3a: spectral analysis and stellar parameters}
\label{sec:mod}

\begin{figure*}
\includegraphics[width=15cm]{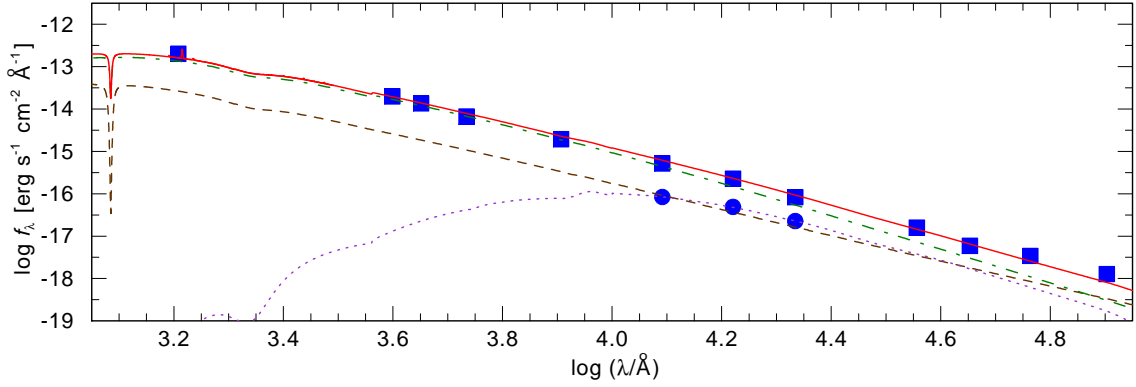}
\centering \caption{Reddened synthetic spectral energy
distributions of the WR (brown dashed line) and O (green
dot-dashed line) components of BAT99\,3a, star\,3 (purple dotted
line), and BAT99\,3a + star\,3 (red solid line). Blue squares
denote the total photometry of BAT99\,3a and star\,3, while blue
circles denote the photometry of star\,3 alone.} \label{fig:sed}
\end{figure*}
\begin{figure*}
\includegraphics[width=15cm,angle=0]{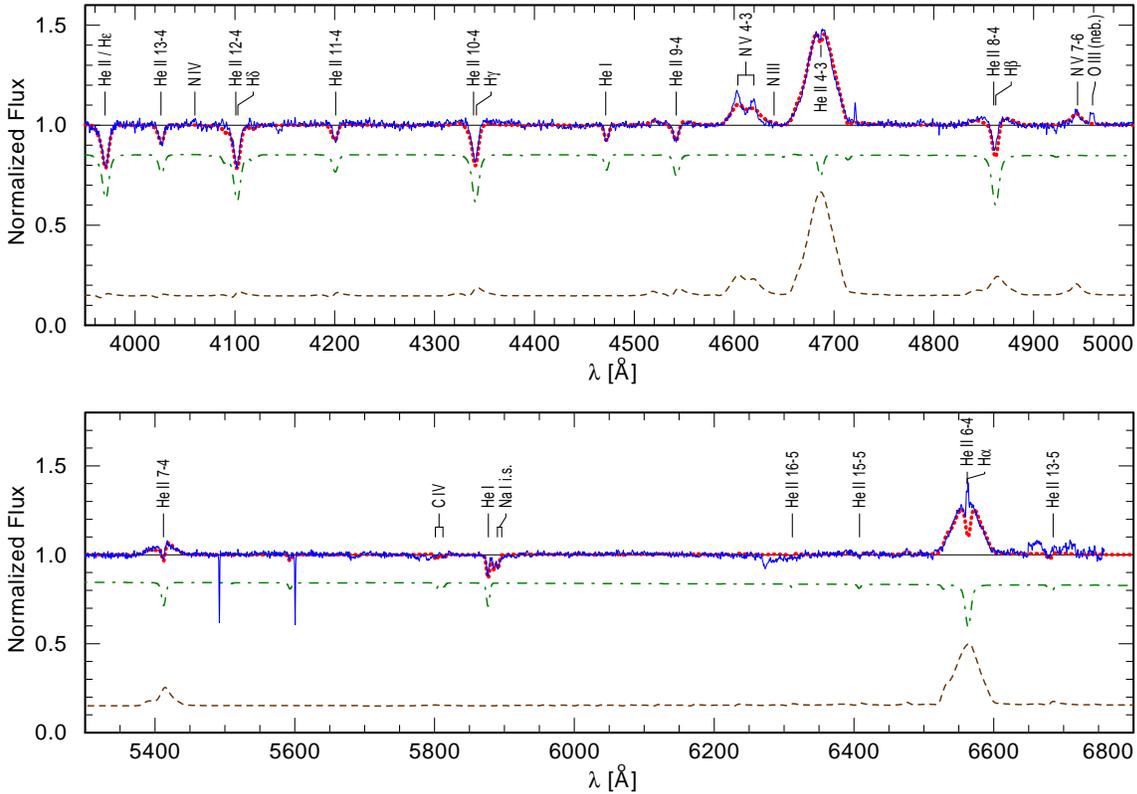}
\centering \caption{Normalized synthetic spectra of the WR (brown
dashed line) and O (green dot-dashed line) components of
BAT99\,3a, as well as of BAT99\,3a + star\,3 (red dotted line).
The component spectra are weighted according to their relative
contributions to the combined flux. Star\,3 hardly contributes to
the flux and is therefore not included. The normalized observed
spectrum, taken at Gemini on 2011 February 9, is also shown (blue
solid line).} \label{fig:fit}
\end{figure*}

The spectral analysis was performed using the state-of-the-art
non-LTE Potsdam Wolf-Rayet (PoWR) code. PoWR is a model atmosphere
code especially suitable for hot stars with expanding atmospheres.
The code solves the radiative transfer and rate equations in the
co-moving frame, calculates non-LTE population numbers, and
delivers a synthetic spectrum in the observer's
frame\footnote{PoWR models of WR stars can be downloaded at
http://www.astro.physik.uni-potsdam.de/PoWR.html}. The
prespecified wind velocity field takes the form of a $\beta$-law
in the supersonic region. In the subsonic region, the velocity
field is defined such that a hydrostatic density stratification is
obtained. A closer description of the assumptions and methods used
in the code is given by Gr\"afener, Koesterke \& Hamann (2002) and
Hamann \& Gr\"afener (2004). The so-called `microclumping'
approach is used to account for wind homogeneities. The density
contrast between a clumped and a smooth wind with an identical
mass-loss rate is described by the clumping factor $D$ (cf. Hamann
\& Koesterke 1998). Line blanketing is treated using the
superlevel approach (Gr\"afener et al. 2002), as originally
introduced by Anderson (1989). We adopt $\beta$=1 for the exponent
in the $\beta$-law. The clumping factor $D$ is set to 10 for the
WR star and to unity for the O star.

A PoWR model atmosphere of a WR star is primarily characterized by
the stellar temperature, $T_\ast$, and the so-called transformed
radius, $R_\mathrm{t}$. $T_\ast$ is the effective temperature of a
star with luminosity $L$ and radius $R_\ast$, as defined by the
Stefan-Boltzmann relation $L=4\pi \sigma R_\ast^2 T_\ast^4$. The
stellar radius $R_\ast$ is defined at the Rosseland optical depth
$\tau_\mathrm{Ross}$=20. $R_\mathrm{t}$ is given by
\begin{eqnarray}
R_{\rm t}=R_\ast \left[\frac{v_\infty}{2500\,{\rm km}\,{\rm
s}^{-1}} \left/ \frac{\dot{M} \sqrt{D}}{10^{-4}\,\msun/ {\rm
yr}}\right.\right]^{2/3} \, , \nonumber
\end{eqnarray}
where $\dot{M}$ is the mass-loss rate and $v_\infty$ is the
terminal wind velocity. For O stars, the effective gravity $\log
g_{\rm eff}$, which is the gravity corrected for the radiative
pressure on free electrons, is used instead of $R_\mathrm{t}$ as a
second fundamental quantity.

The model parameters are determined by iterative fitting the
spectral energy distribution (SED) and synthetic normalized
spectrum to the observations (see Figs\,\ref{fig:sed} and
\ref{fig:fit}, respectively). The synthetic composite spectrum
used for BAT99\,3a is the sum of two models corresponding to its
two components, weighted according to their relative contribution
to the overall flux. The temperatures are obtained from the line
ratios of different ions, while the mass-loss rate of the WR
component is derived from the strengths of emission lines. The
effective gravity $\log g_{\rm eff}$ of the O companion is
inferred from the wings of prominent hydrogen lines. The total
luminosity $L_1 + L_2$ is determined from the observed SED, while
the light ratio is obtained from the normalized spectrum. In our
case, the effect of the light ratio is most easily isolated from
the other parameters by simultaneously inspecting lines which
clearly originate in one of the components, e.g. He\,{\sc i} lines
belonging to the O component, or N\,{\sc v} and He\,{\sc ii}
emission lines from the WR component. Once the light ratio and
total luminosity are known, the individual luminosities follow.

Fig.\,\ref{fig:sed} plots the synthetic SEDs of the best-fit
models for both components of BAT99\,3a as well as for star\,3.
The total synthetic SED belonging to BAT99\,3a and star\,3 (red
solid line) is also plotted. The blue squares denote the total
far-ultraviolet (FUV) (Parker et al. 1998), UBVI (Zaritsky et al.
2004), $JHK_{\rm s}$ (Kato et al. 2007) and IRAC photometry of
BAT99\,3a and star\,3. The FUV magnitude also includes a
contribution from the nearby B star (star\,2). This contribution,
however, is insignificant because of the spectral type of the
star. On the other hand, star\,2 might somewhat influence the IRAC
5.8 and 8 $\mu$m photometry, which is indeed implied by the small
observed IR excess compared to the synthetic SED in
Fig.\,\ref{fig:sed}. Since the $JHK_{\rm s}$ survey by Kato et al.
(2007) resolves BAT99\,3a and star\,3, we denote the $JHK_{\rm s}$
magnitudes corresponding to star\,3 only (blue circles) as well.
The corresponding normalized spectra of the different components
are shown in Fig.\,\ref{fig:fit}. The contribution of star\,3 is
negligible and therefore not shown. The sum of all model spectra
(red dotted line) is further compared with the Gemini spectrum
observed on 2011 February 9 (blue solid line).

The stellar parameters and abundances used for the best-fitting WR
and O star models are compiled in Table\,\ref{tab:model}, where we
also provide the number of ionizing photons per second for
hydrogen ($Q_0$), He\,{\sc i} ($Q_1$) and He\,{\sc ii} ($Q_2$).
For the O star we also give the gravity, $\log g$, and a projected
rotational velocity, $v\sin i$. The absolute visual magnitudes
inferred for the binary components imply a $V$-band magnitude
difference of $\Delta V=1.89$ mag, i.e. the O component
contributes $\approx$5.7 times more than the WR component in the
visual band. The H and N abundances for the WR star are derived
from the spectral analysis. The adopted abundances of the
remaining elements, which cannot be derived from the available
spectra, are based on a recent work by Hainich et al. (2014). For
the O component, we adopt typical LMC abundances (Hunter et al.
2007; Trundle et al. 2007).

For the WR component, we also estimate the initial mass, $M_{\rm
i}$, and age of $\approx$30\,$\msun$ and $\approx$7 Myr,
respectively, using the single star evolutionary tracks from
Meynet \& Maeder (2005). To estimate the current mass of the WR
component, $M_{\rm cur}$, we make use of the mass-luminosity
relation for homogenous WR stars (Gr\"{a}fener et al. 2011; see
their equation\,11), which is dependent of the hydrogen abundance.
If the WR star is indeed homogenous, e.g. because it is a rapid
rotator (Heger \& Langer 2000), our analysis implies a hydrogen
abundance in the stellar core of $\approx$0.2, and the resulting
$M_{\rm cur}$ is $23 \, \msun$. More likely, however, the core is
almost depleted of hydrogen, which would imply that $M_{\rm cur}$
is $14 \, \msun$. Since rotational velocities of WR stars are
poorly understood at present and were claimed to be indirectly
observed only in a few WR stars (Chen\'{e} \& St-Louis 2005,
Shenar et al. 2014), we may merely conclude that $M_{\rm cur}$
lies somewhere between these two extremes. The relatively large RV
amplitude of the WR component in comparison with the primary O
star (see Table\,\ref{tab:RVs} and Fig.\,\ref{fig:RVc}), however,
suggests a value closer to $14\,M_\odot$ than to $23\,M_\odot$.

On the other hand, the single star evolutionary tracks from
K\"{o}hler et al. (in preparation)\footnote{See also
http://bonnsai.astro.uni-bonn.de} give for the O star an initial
mass and age of $\approx$$30 \, \msun$ and $\approx$3 Myr,
respectively. The latter figure is much smaller than the estimated
age of the WR star of 7 Myr, which implies that there was mass
transfer in the binary system and that the O star is a rejuvenated
mass gainer (e.g. Schneider et al. 2014), and that $M_{\rm i}$ of
the WR star cannot be estimated from single star models. For
$M_{\rm cur} =14 \, \msun$, one can expect that $M_{\rm i}$ of the
WR star was about $40 \, \msun$ (Wellstein \& Langer 1999), which
would imply an age of this star of $\approx$6 Myr. What is
unexpected in the above consideration is to have a large
eccentricity of the binary system after the mass transfer. This
could, however, be understood if the system was kicked out of the
parent star cluster (after the mass transfer was completed) either
because of dynamical encounter with another massive (binary)
system or due to dissolution of a multiple system (cf.
Section\,\ref{sec:rel}).

\begin{table}
\caption{Stellar parameters for the WN3 and O-type components of
BAT99\,3a.} \label{tab:model}
\begin{center}
\begin{tabular}{lrr}
\hline
 & WN3 & O6\,V \\
\hline
$T_{\ast}$ [kK] & 89 & 38 \\
$\log R_{t}$ [$\rsun$] & 1.0 & --- \\
$\log g$ [cm\,s$^{-2}$] & --- & 3.73 \\
$\log g_{\rm eff}$ [cm\,${\rm s}^{-2}$] & --- & 3.60 \\
$\log L$ [$\lsun$] & 5.45 & 5.20 \\
$v_{\infty}$ [$\kms$] & 1800 & 2500 \\
$D$ & 10 & 1 \\
$R_{\ast}$ $\lbrack \rsun \rbrack$ & 2.23 & 9.20 \\
$\log \dot{M}$ [$\myr$] & $-$5.8 & $-$7.0 \\
$\log Q_0$ [s$^{-1}$] & 49.4 & 48.82 \\
$\log Q_1$ [s$^{-1}$] & 49.2 & 47.79 \\
$\log Q_2$ [s$^{-1}$] & 46.2 & 42.31 \\
$v \sin i$ [km\,s$^{-1}$] & --- & 80 \\
$E(B-V)$ [mag] & 0.18 & 0.18 \\
$A_V$ [mag] & 0.56 & 0.56 \\
$R_V$ & 3.1 & 3.1 \\
$M_V$ [mag] & $-$2.86 & $-$4.75 \\
H (mass fraction) & 0.20 & 0.73 \\
He (mass fraction) & 0.80 & 0.27 \\
C (mass fraction) & $0.7\times10^{-4}$ & $4.7\times10^{-4}$ \\
N (mass fraction) & 0.008 & $7.8\times10^{-5}$ \\
Fe (mass fraction) & $7.0\times10^{-4}$ & $7.0\times10^{-4}$ \\
\hline
\end{tabular}
\end{center}
\end{table}

While it is not possible to quantify the detailed stellar
parameters of star\,3 with the current data, the photometric data
imply that this star is highly reddened. With no indications for a
later-type star in the available spectrum, we conclude that
star\,3 is of O or B type. In Fig.\,\ref{fig:sed} we illustrate
that the observed photometry is consistent with the SED typical of
an O8.5\,V star with an adopted colour excess $E(B-V)$=1.75 mag
(cf. Sect.\,\ref{sec:rel}).

Since the O component of BAT99\,3a does not show clear evidence
for wind lines in the available spectra, a spectroscopic
calculation of its $v_\infty$ and $\dot{M}$ is not possible.
Nevertheless, $\dot{M}$ can be constrained because larger values
would lead to observable effects on the spectrum. The value of
$\dot{M}$ given in Table\,\ref{tab:model} for the O component
should therefore be considered as an upper limit. As no
information regarding $v_\infty$ of the O component is available
from its spectrum, we merely adopt a value typical of stars of
similar spectral type (see Table\,\ref{tab:model}), based on the
work by Mokiem et al. (2007).

For the Galactic interstellar reddening, we adopt a colour excess
of $E(B-V)$=0.03 mag. To model the LMC interstellar reddening, we
assume the reddening law suggested by Howarth (1983) with
$R_V$=3.1. The best fit to the whole SED was achieved with
$E(B-V)$=0.18 mag, which is comparable to that of 0.22 mag derived
for star\,2 in Section\,\ref{sec:B}. On the other hand, a similar
fit to the SED could also be obtained with a colour excess of 0.12
mag if the luminosities of the two companions are both reduced by
0.1\,dex. This $E(B-V)$ is closer to the one obtained from the
analysis of the circular shell (see Section\,\ref{sec:neb}).
Future UV observations should enable a much more precise
determination of the luminosities and redenning.

A comparison of the stellar parameters of the WR component with
the almost complete sample of the WR stars in the LMC (Hainich et
al. 2014) reveals that the analyzed WR star is one of the hottest
WR stars in the LMC. While $Q_0$ and $Q_1$ of the WR component are
comparable to those of its O companion, the number of He\,{\sc ii}
ionizing photons is four orders of magnitude larger, placing
BAT99\,3a among the strongest He\,{\sc ii} ionizing sources in the
LMC. This is in very good agreement with the unique signature of
He\,{\sc ii} emission observed in the circular shell, as discussed
in Section\,\ref{sec:neb}.

\section{Circular shell}
\label{sec:neb}

\subsection{Circular shells around WR stars}

Johnson \& Hogg (1965) were the first who noted association of WR
stars with filamentary shells and suggested that they are the
result of interaction between the material ejected from the stars
and the interstellar matter (cf. Avedisova 1972). This suggestion
was further elaborated to take into account the interaction of the
WR wind with the circumstellar material shed during the preceding
red supergiant or LBV stages (e.g. D'Ercole 1992;
Garc\'{i}a-Segura, Langer \& Mac Low 1996a,b; Brighenti \&
D'Ercole 1997). The {\it circumstellar} shells produced in this
process show signatures of CNO-processing (e.g. Esteban et al.
1992; Stock, Barlow \& Wesson 2011) and are associated exclusively
with WNL stars (e.g. Lozinskaya \& Tutukov 1981; Chu 1981; Gruendl
et al. 2000; Gvaramadze et al. 2010a), i.e. the very young WR
stars, whose winds are still confined within the region occupied
by the circumstellar material or only recently emerged from it
(Gvaramadze et al. 2009; Burgemeister et al. 2013). The size and
crossing time of this region determine the characteristic size and
lifetime of the circumstellar shells, which are typically several
pc and several tens of thousands of years, respectively. The winds
of more evolved (WNE) WR stars interact directly with the local
interstellar medium and create new, generally more extended
shells. Correspondingly, the chemical composition of these {\it
interstellar} shells is similar to that of ordinary \hii regions
(e.g. Esteban et al. 1992; Stock et al. 2011).

Since the WR component of BAT99\,3a is of WNE type, it is natural
to expect that its associated shell should consist mostly of
interstellar matter. In Section\,\ref{sec:neb-spec}, we support
this expectation by an analysis of the long-slit spectra of the
shell obtained with the Gemini and du Pont telescopes and the
SALT.

\begin{figure}
\begin{center}
\includegraphics[width=7.5cm,angle=0]{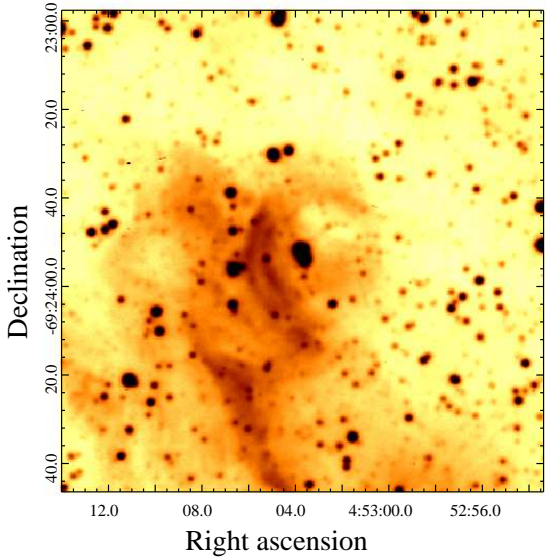}
\end{center}
\caption{MCELS2 H$\alpha$ image of the circular shell. BAT99\,3a
and star\,2 are clearly resolved in the image. At the distance of
the LMC of 50 kpc, 20 arcsec corresponds to $\approx$4.8 pc.}
\label{fig:c&g}
\end{figure}

\subsection{H$\alpha$ imaging of the shell}
\label{sec:mul}

Before discussing the long-slit spectra, we present an H$\alpha$
image of the shell obtained as part of the Magellanic Clouds
Emission Line Survey 2 (MCELS2; PI: Y.-H. Chu). This survey used
the MOSAIC\,{\sc ii} camera on the Blanco 4-m telescope at the
Cerro Tololo Inter-American Observatory (Muller et al. 1998) to
image the entire LMC and SMC in the H-alpha line only. The region
of the LMC containing the shell was observed on 2008 December 12
with three dithered 300 s exposures, through a 80\,\AA \,
bandwidth filter centred at 6563\,\AA. The image was overscan,
crosstalk, bias, and flat-field corrected using standard routines
in {\sc iraf}. An astrometric solution and distortion correction
were obtained by comparison to 2MASS Point Source Catalog
(Skrutskie et al. 2006), and have a typical accuracy of
$\approx$150 mas.

The H$\alpha$ image of the shell is presented in
Fig.\,\ref{fig:c&g}. The shell appears as an incomplete circle
bounded on the eastern side by two concentric arc-like filaments,
separated from each other by $\approx$4 arcsec (or $\approx$1 pc),
and on the northern side by a thick amorphous arc. There is no
distinct limb-brightened emission in the southwest direction, but
the shell can still be discerned because of a diffuse emission
filling its interior. BAT99\,3a and star\,2 are clearly resolved
in the image. The lopsided appearance of the shell and the
off-centred position of the stars suggest that the ambient medium
is denser in the eastern direction, which is consistent with the
presence of a prominent \hii region in this direction (cf.
Section\,\ref{sec:nebula} and see also the next section).

\subsection{Spectroscopy of the shell}
\label{sec:neb-spec}

All long-slit spectra obtained with the Gemini and du Pont
telescopes and the SALT show numerous emission lines along the
whole slit. Intensities of these lines depend on weather
conditions and the slit's PA. Some of the lines show spatial
correlation with the shell. Below we analyse the obtained spectra
to understand the origin of the shell.

\begin{figure*}
 \includegraphics[angle=-90,width=15cm,clip=]{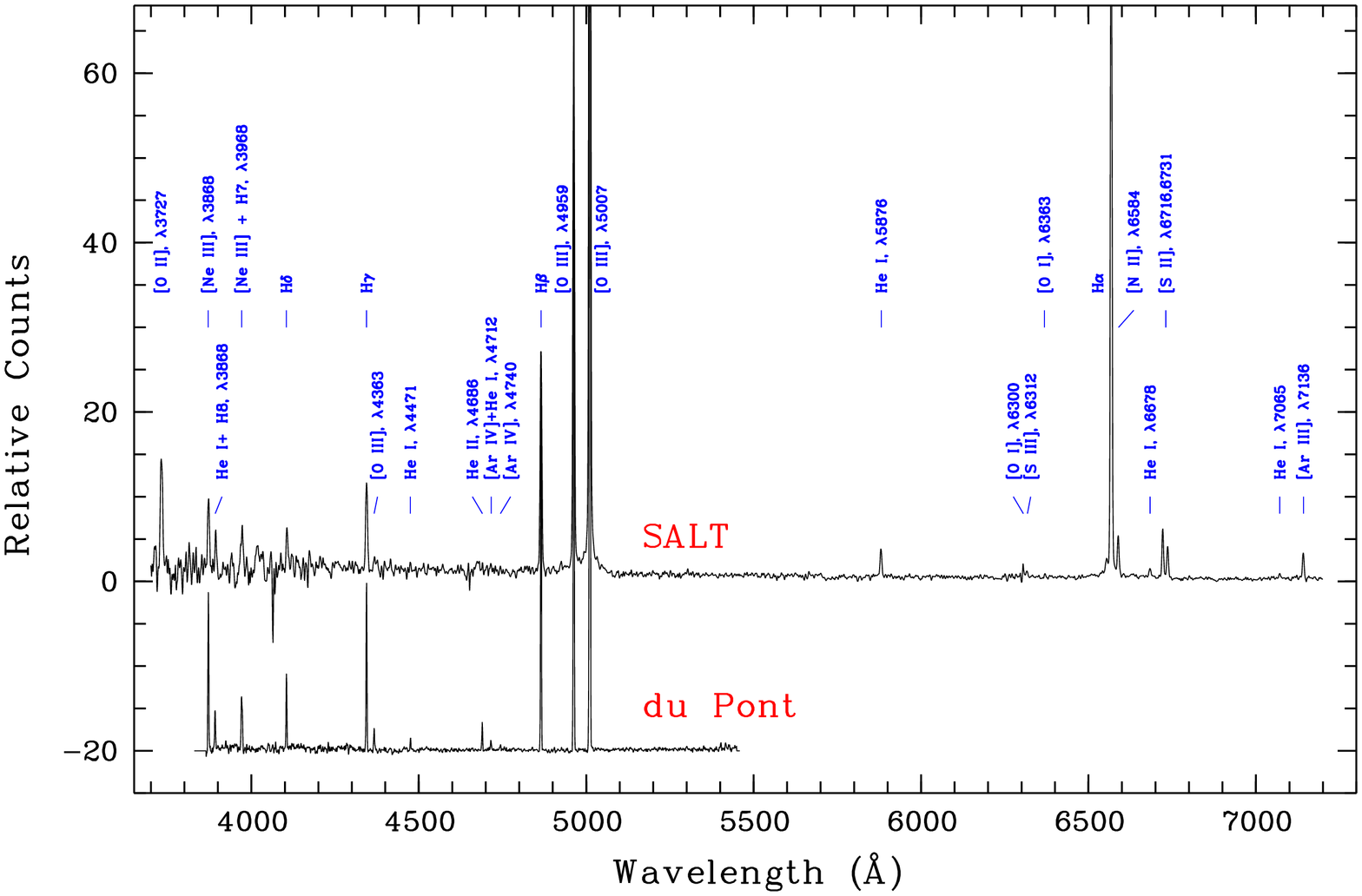}
 \caption{1D reduced spectra of the circular shell obtained with
  the SALT and du Pont telescopes (the du Pont spectrum is
  shifted downwards by 20 counts). All detected emission lines
  are marked.}
 \label{fig:Neb_1d_spec}
\end{figure*}

\subsubsection{Long-slit spectra of the shell}
\label{sec:reduction}

For the analysis of physical conditions and elemental abundances
in the shell all long-slit spectra were re-reduced in the same
manner and corrected for the night-sky background and emission
from the fore/background \hii region using spectral data outside
the area of radius of 25 arcsec centred on BAT99\,3a. The finally
reduced SALT relative flux distribution for the WN star was used
to construct the sensitivity curve for the du Pont data. Since the
spectra were obtained under different weather conditions, only
some of them show diagnostic lines with a good enough S/N ratio.
For this reason, only the first Gemini, the second du Pont and
both SALT spectra (see Table\,\ref{tab:log}) were used in our
following analysis.

One-dimensional (1D) spectra of the shell were extracted from the
area of radius of 15 arcsec centred on BAT99\,3a with the central
$\pm$3 arcsec excluded to avoid the effect of BAT99\,3a and
star\,2. The final 1D du Pont and SALT spectra with all identified
emission lines are presented in Fig.\,\ref{fig:Neb_1d_spec}. We do
not show the Gemini spectrum because of its much worse quality
caused by poor weather conditions during the observing run.

All spectra show the nebular He\,{\sc ii} $\lambda$4686 emission
associated with the shell (see below). Only few such
high-excitation nebulae are known in the Local Group (e.g. Garnett
et al. 1991; Naz\'{e} et al. 2003; Pakull 2009; Kehrig et al.
2011) and most of them are associated with WNE, WC and WO stars,
i.e. with WR stars capable of ionizing He\,{\sc ii}.

Emission lines in the spectra of the shell were measured using
programs described in detail in Kniazev et al. (2004, 2005). These
programs determine the location of the continuum, perform a robust
noise estimation, and fit separate lines by a single Gaussian
superimposed on the continuum-subtracted spectrum. Some
overlapping lines were fitted simultaneously as a blend of two or
more Gaussian features: the H$\alpha$ $\lambda$6563 and [N\,{\sc
ii}] $\lambda\lambda$6548, 6584 lines, the [S\,{\sc ii}]
$\lambda\lambda$6716, 6731 lines, and the [O\,{\sc i}]
$\lambda$6300 and [S\,{\sc iii}] $\lambda$6312 lines.

%-----------------------------------------------------------------------------
\begin{table*}
\begin{center}
\caption{Line intensities of the shell.} \label{tab:int3}
\begin{tabular}{lccccc} \hline
\rule{0pt}{10pt}
              & \MC{1}{c}{Gemini}  & \MC{1}{c}{du Pont} & \MC{1}{c}{Gemini+du Pont} & \MC{2}{c}{SALT} \\
\rule{0pt}{10pt} $\lambda_{0}$(\AA) Ion             &
$F(\lambda)/F$(H$\beta$) & $F(\lambda)/F$(H$\beta$) &
$I(\lambda)/I$(H$\beta$) & $F(\lambda)/F$(H$\beta$) & $I(\lambda)/I$(H$\beta$) \\
\hline
3727\ [O\ {\sc ii}]\              & ---             &~0.755$\pm$0.055$^a$ & 0.839$\pm$0.064 &  0.755$\pm$0.055 & 0.817$\pm$0.061   \\
3868\ [Ne\ {\sc iii}]\            & ---             & 0.420$\pm$0.011     & 0.458$\pm$0.012 &  0.396$\pm$0.024 & 0.423$\pm$0.026   \\
3967\ [Ne\ {\sc iii}]+H7\         & ---             & 0.222$\pm$0.013     & 0.240$\pm$0.015 &  ---             & ---               \\
4101\ H$\delta$\                  & ---             & 0.194$\pm$0.006     & 0.207$\pm$0.007 &  0.215$\pm$0.029 & 0.226$\pm$0.031   \\
4340\ H$\gamma$\                  & 0.508$\pm$0.029 & 0.453$\pm$0.011     & 0.472$\pm$0.012 &  0.420$\pm$0.017 & 0.434$\pm$0.017   \\
4363\ [O\ {\sc iii}]\             & 0.066$\pm$0.017 & 0.069$\pm$0.004     & 0.072$\pm$0.005 &  0.059$\pm$0.009 & 0.061$\pm$0.009   \\
4471\ He\ {\sc i}\                & ---             & 0.030$\pm$0.002     & 0.031$\pm$0.002 &  0.029$\pm$0.011 & 0.029$\pm$0.011   \\
4686\ He\ {\sc ii}\               & 0.071$\pm$0.014 & 0.071$\pm$0.004     & 0.072$\pm$0.004 &  0.062$\pm$0.019 & 0.062$\pm$0.020   \\
4712\ [Ar\ {\sc iv]}+He\ {\sc i}  & ---             & 0.026$\pm$0.003     & 0.026$\pm$0.003 &  ---             & ---               \\
4861\ H$\beta$\                   & 1.000$\pm$0.039 & 1.000$\pm$0.030     & 1.000$\pm$0.030 &  1.000$\pm$0.043 & 1.000$\pm$0.043   \\
4959\ [O\ {\sc iii}]\             & 2.449$\pm$0.076 & 2.423$\pm$0.055     & 2.406$\pm$0.055 &  2.317$\pm$0.080 & 2.304$\pm$0.079   \\
5007\ [O\ {\sc iii}]\             & 7.221$\pm$0.219 & 7.251$\pm$0.163     & 7.172$\pm$0.161 &  6.771$\pm$0.213 & 6.713$\pm$0.212   \\
5876\ He\ {\sc i}\                & 0.114$\pm$0.012 & ---                 & 0.106$\pm$0.011 &  0.133$\pm$0.018 & 0.126$\pm$0.018   \\
6300\ [O\ {\sc i}]\               & ---             & ---                 & ---             &  0.035$\pm$0.005 & 0.033$\pm$0.005   \\
6312\ [S\ {\sc iii}]\             & ---             & ---                 & ---             &  0.020$\pm$0.005 & 0.019$\pm$0.004   \\
6548\ [N\ {\sc ii}]\              & 0.046$\pm$0.010 & ---                 & 0.042$\pm$0.009 &  0.054$\pm$0.004 & 0.049$\pm$0.004   \\
6563\ H$\alpha$\                  & 3.170$\pm$0.092 & ---                 & 2.859$\pm$0.090 &  3.102$\pm$0.097 & 2.852$\pm$0.097   \\
6584\ [N\ {\sc ii}]\              & 0.148$\pm$0.013 & ---                 & 0.134$\pm$0.012 &  0.163$\pm$0.009 & 0.149$\pm$0.008   \\
6678\ He\ {\sc i}\                & ---             & ---                 & ---             &  0.060$\pm$0.025 & 0.055$\pm$0.023   \\
6716\ [S\ {\sc ii}]\              & 0.264$\pm$0.021 & ---                 & 0.237$\pm$0.019 &  0.194$\pm$0.008 & 0.177$\pm$0.007   \\
6731\ [S\ {\sc ii}]\              & 0.200$\pm$0.019 & ---                 & 0.179$\pm$0.017 &  0.142$\pm$0.006 & 0.129$\pm$0.006   \\
7065\ He\ {\sc i}\                & ---             & ---                 & ---             &  0.019$\pm$0.012 & 0.017$\pm$0.011   \\
7136\ [Ar\ {\sc iii}]\            & ---             & ---                 & ---             &  0.112$\pm$0.017 & 0.100$\pm$0.015   \\
  & & \\
$C$(H$\beta$)\ dex        &&& \MC {1}{c}{0.14$\pm$0.04} & \MC {2}{c}{0.11$\pm$0.04} \\
$E(B-V)$\ mag             &&& \MC {1}{c}{0.10$\pm$0.03} & \MC {2}{c}{0.08$\pm$0.04} \\
\hline
\multicolumn{4}{l}{$^a$Based on the SALT data.}\\
\end{tabular}
\end{center}
\end{table*}
%-----------------------------------------------------------------------------

Table\,\ref{tab:int3} lists the observed intensities of all
detected lines normalized to H$\beta$, F($\lambda$)/F(H$\beta$).
To have independent estimates of the shell parameters, we combined
the Gemini and du Pont data and analysed them separately from the
SALT ones. The relative intensities of lines detected both in the
Gemini and du Pont spectra, e.g. H$\gamma$, [O\,{\sc iii}]
$\lambda$4363 and [O\,{\sc iii}] $\lambda\lambda$4959, 5007, are
consistent with each other within the individual rms uncertainties
(see columns 2 and 3 in Table\,\ref{tab:int3}) so that in the
analysis based on the Gemini+du Pont data, we used the blue
($<$5876 \AA) lines from the du Pont spectrum only and the red
lines from the Gemini one. The combined SALT data cover the total
spectral range 3700$-$7200\,\AA. The bluest ($<$4200~\AA) and
reddest ($>$6700~\AA) lines in this data set have the largest
relative errors because these lines were covered by only one
observation (see Table\,\ref{tab:log}). Since the Gemini+du Pont
data set does not contain any of the O$^{+}$ lines required to
derive the O$^{+}$/H$^{+}$ abundance, we added to these data the
intensity of the [O\,{\sc ii}] $\lambda$3727 line from the
combined SALT spectrum (see Table\,\ref{tab:int3}).

Table\,\ref{tab:int3} also lists the reddening-corrected line
intensity ratios, I($\lambda$)/I(H$\beta$), and the logarithmic
extinction coefficient, $C$(H$\beta$). The extinction coefficients
derived from both sets of data (Gemini+du Pont and SALT) are
consistent with each other and correspond to $E(B-V)$=0.08$-$0.10
mag, which is a factor of two less than the colour excess derived
for BAT99\,3a and star\,2 from the spectral analysis and
photometry, respectively (see Sections\,\ref{sec:mod} and
\ref{sec:B}). Since the spectra of the shell do not show any
significant continuum (see Fig.\,\ref{fig:Neb_1d_spec}), we
assumed in our calculations that EW of underlying absorption in
Balmer hydrogen lines is =0 \AA \, (cf. Kniazev et al. 2004,
2005).

\subsubsection{Physical conditions and elemental abundances}
\label{sec:phys}

To estimate the parameters of the shell, we used the technique of
plasma diagnostics in the way described in full detail in Kniazev
et al. (2008b). The electron temperature, $T_{\rm e}$([O\,{\sc
iii}]), was calculated using the weak auroral line of oxygen
[O\,{\sc iii}] $\lambda$4363, while the electron number density,
$n_{\rm e}$([S\,{\sc ii}]), was derived from the intensity ratio
of the [S\,{\sc ii}] $\lambda\lambda$6716, 6731
lines\footnote{Although the [Ar\,{\sc iv}] $\lambda\lambda$4711,
4740 doublet is also present in the du Pont spectrum of the shell,
we did not use it for density evaluation because of its
weakness.}. The estimates of $n_{\rm e}$ based on the two data
sets are consistent with each other and are typical of \hii
regions and interstellar shells around WR stars (e.g. Esteban et
al. 1992; Stock et al. 2011).

From the observed emission lines, we determined the abundances of
O, N and Ne using the Gemini+du Pont data and those of O, N, Ne,
S, Ar and He using the SALT data. As described in Kniazev et al.
(2008b), we calculate ionic abundances, the ionization correction
factors (ICFs) and total element abundances using system of
equations from Izotov et al. (2006), which are based on sequences
of photoionization models and used the new atomic data of
Stasi\'nska (2005). These abundances along with directly
calculated $T_{\rm e}$, $n_{\rm e}$, ICFs for different elements,
and total elemental abundances are listed in
Table\,\ref{tab:Chem}, where we also give the LMC abundances from
Russel \& Dopita (1992). Table\,\ref{tab:Chem} also shows $T_{\rm
e}$([O\,{\sc ii}]) and $T_{\rm e}$([S\,{\sc iii}]), calculated
using approximations from Izotov et al. (2006). We used $T_{\rm
e}$([S\,{\sc iii}]) for the calculation of S$^{2+}$/H$^+$ and
Ar$^{2+}$/H$^+$ abundances. The detection of a strong nebular
He\,{\sc ii} $\lambda$4686 emission implies the presence of a
non-negligible amount of O$^{3+}$, whose abundance can be derived
in the way described in Izotov et al. (2006).

The helium abundance was derived in the manner described in detail
in Izotov \& Thuan (1994), Izotov, Thuan \& Lipovetsky (1994) and
Izotov, Thuan \& Lipovetsky (1997). The new fits from Benjamin,
Skillman \& Smith (2002) were used to convert He\,{\sc i} emission
line strengths to singly ionized helium abundances.

One can see that the shell abundances derived from both data sets
are consistent with each other and the LMC ones within the error
margins, which implies the absence of N and He enrichment of the
shell and suggests that it is mainly composed of material swept-up
from the local interstellar medium.

\begin{table*}
\centering{ \caption{Elemental abundances in the shell.}
\label{tab:Chem}
\begin{tabular}{lccc} \hline
\rule{0pt}{10pt} Quantity               & Gemini+du Pont        &
SALT            & LMC$^a$     \\ \hline
$T_{\rm e}$([O\,{\sc iii}])(K)\         & 11,600$\pm$300~~      & 11,200$\pm$600~~   &                  \\
$T_{\rm e}$([O\,{\sc ii}])(K)\          & 12,000$\pm$200~~      & 11,400$\pm$200~~   &                  \\
$T_{\rm e}$([S\,{\sc iii}])(K)\         & ---                   & 11,300$\pm$760~~   &                  \\
$n_{\rm e}$([S\,{\sc ii}])(cm$^{-3}$)\  & 90$^{+200}_{-80}$~~   & 50$^{+90}_{-40}$~~ &                  \\
& \\
O$^{+}$/H$^{+}$($\times$10$^5$)\        & 1.615$\pm$0.144~~     & 1.942$\pm$0.189~~  &                  \\
O$^{++}$/H$^{+}$($\times$10$^5$)\       & 16.200$\pm$1.211~~    & 17.060$\pm$2.812~~ &                  \\
O$^{+++}$/H$^{+}$($\times$10$^5$)\      & 1.349$\pm$0.178~~     & 0.696$\pm$0.377~~  &                  \\
O/H($\times$10$^5$)\                    & 19.160$\pm$1.779~~    & 19.700$\pm$2.843~~ &                  \\
12+log(O/H)\                            & ~8.28$\pm$0.03~~      & ~8.29$\pm$0.06~~   &   8.35$\pm$0.06  \\
& \\
N$^{+}$/H$^{+}$($\times$10$^7$)\        & 15.870$\pm$1.382~~    & 20.610$\pm$1.253~~ &                  \\
ICF(N)\                                 & 9.789                 & 8.583              &                  \\
N/H($\times$10$^5$)\                    & 1.55$\pm$0.14~~       & 1.77$\pm$0.11~~    &                  \\
12+log(N/H)\                            & 7.19$\pm$0.04~~       & 7.25$\pm$0.06~~    &  7.14$\pm$0.15   \\
%log(N/O)\                               & $-$1.09$\pm$0.05~~    & $-$1.05$\pm$0.07~~ &                  \\
& \\
Ne$^{++}$/H$^{+}$($\times$10$^5$)\      & 2.841$\pm$0.256~~     & 2.999$\pm$0.608~~  &                  \\
ICF(Ne)\                                & 1.073                 & 1.046              &                  \\
Ne/H($\times$10$^5$)\                   & 3.05$\pm$0.28~~       & 3.14$\pm$0.64~~    &                  \\
12+log(Ne/H)\                           & 7.48$\pm$0.05~~       & 7.50$\pm$0.09~~    &  7.61$\pm$0.05   \\
& \\
S$^{+}$/H$^{+}$($\times$10$^7$)\        & ---                   & 5.119$\pm$0.245~~  &                  \\
S$^{++}$/H$^{+}$($\times$10$^7$)\       & ---                   & 25.950$\pm$9.201~~ &                  \\
ICF(S)\                                 & ---                   & 2.179              &                  \\
S/H($\times$10$^7$)\                    & ---                   & 67.71$\pm$20.06~~  &                  \\
12+log(S/H)\                            & ---                   & 6.83$\pm$0.13~~    &  6.70$\pm$0.09   \\
& \\
Ar$^{++}$/H$^{+}$($\times$10$^7$)\      & ---                   & 6.952$\pm$1.466~~  &                  \\
ICF(Ar)\                                & ---                   & 1.240              &                  \\
Ar/H($\times$10$^7$)\                   & ---                   & 8.62$\pm$1.82~~    &                  \\
12+log(Ar/H)\                           & ---                   & 5.94$\pm$0.09~~    &  6.29$\pm$0.25   \\
& \\
12+log(He/H)\                           & ---                   & 10.97$\pm$0.05~~   &  10.94$\pm$0.03  \\
\hline \multicolumn{3}{l}{$^a$The LMC abundances are from Russell
\& Dopita (1992).}
\end{tabular}
 }
\end{table*}

\subsubsection{Position--velocity diagrams along the slit}
\label{sec:PV}

\begin{figure*}%[H]
\begin{minipage}[h]{0.49\linewidth}
\center{\includegraphics[width=1\linewidth]{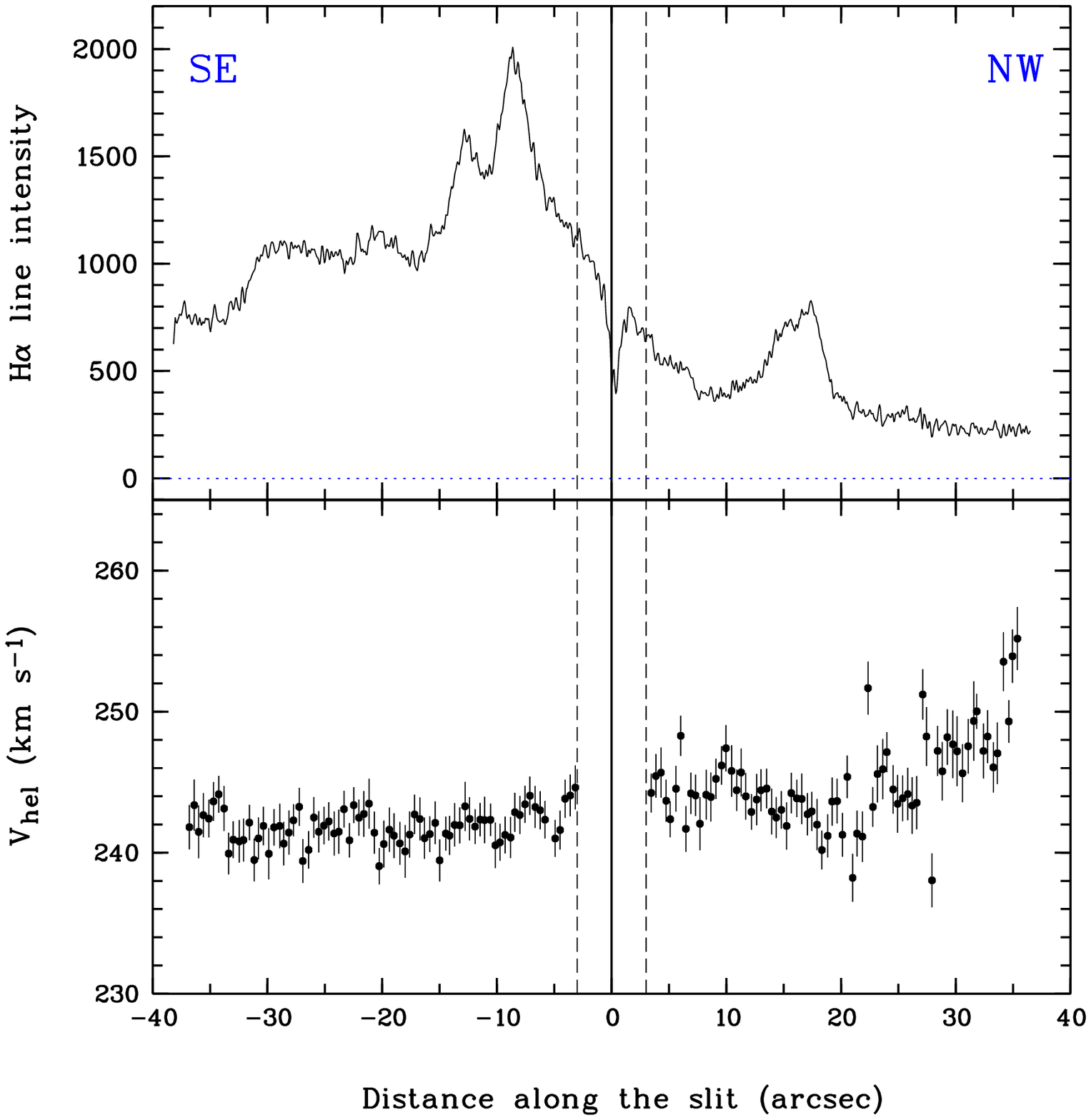}} {} \\
\end{minipage}
\hfill
\begin{minipage}[h]{0.49\linewidth}
\center{\includegraphics[width=1\linewidth]{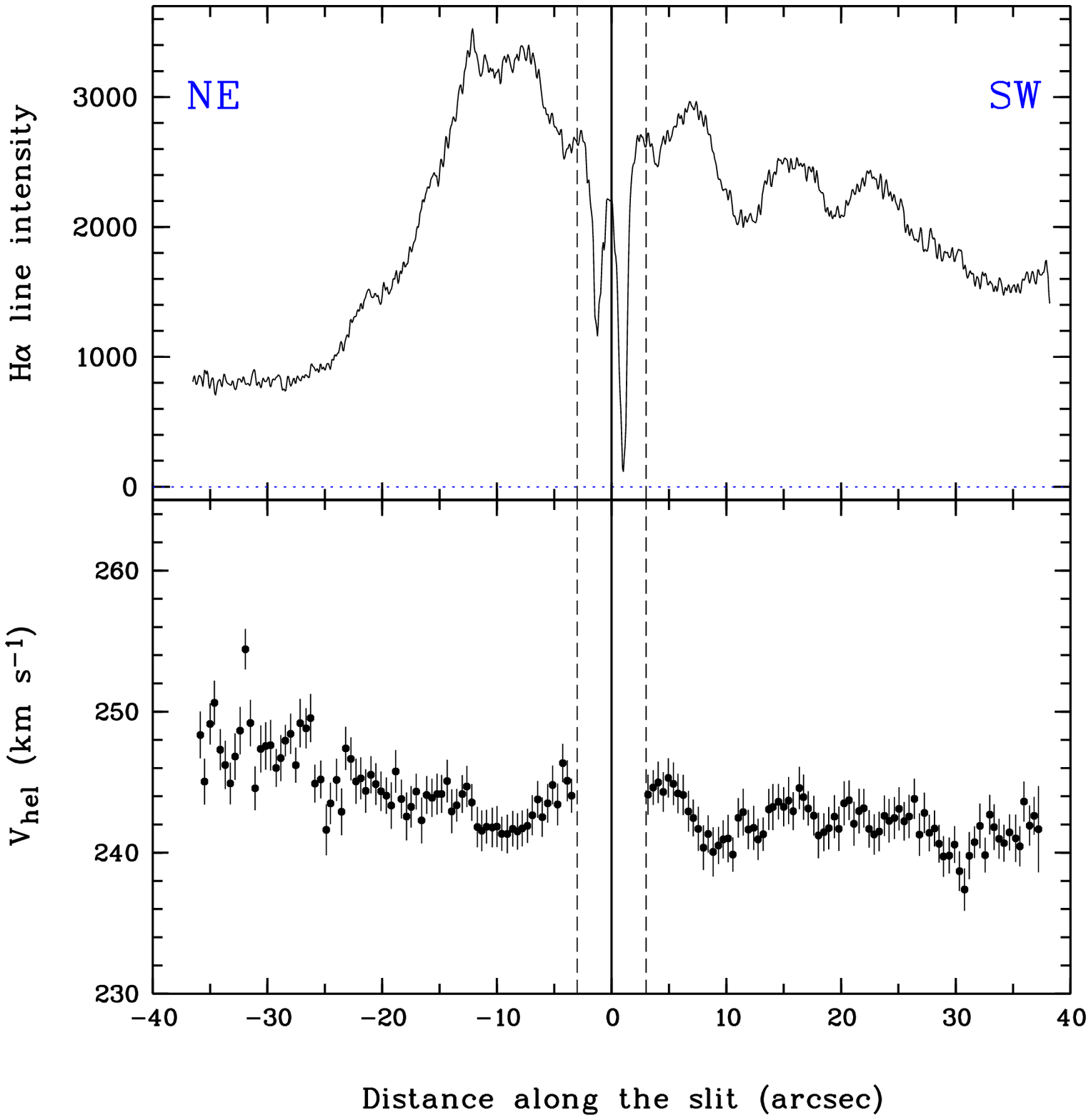}} {} \\
\end{minipage}
\vfill
\begin{minipage}[h]{0.49\linewidth}
\center{\includegraphics[width=0.88\linewidth,angle=270,clip=0]{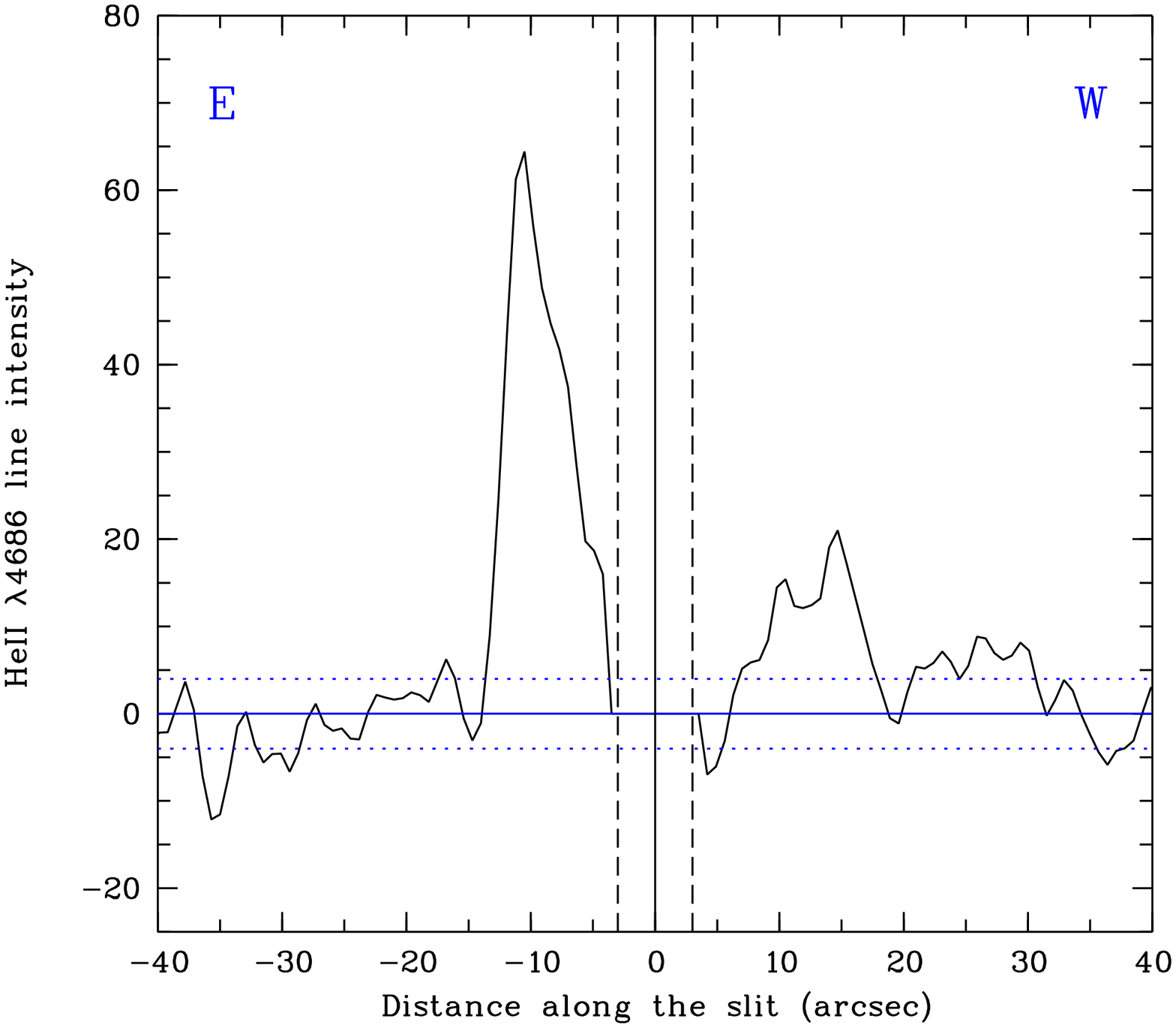}} {} \\
\end{minipage}
\hfill
\begin{minipage}[h]{0.45\linewidth}
\center{\includegraphics[width=0.95\linewidth]{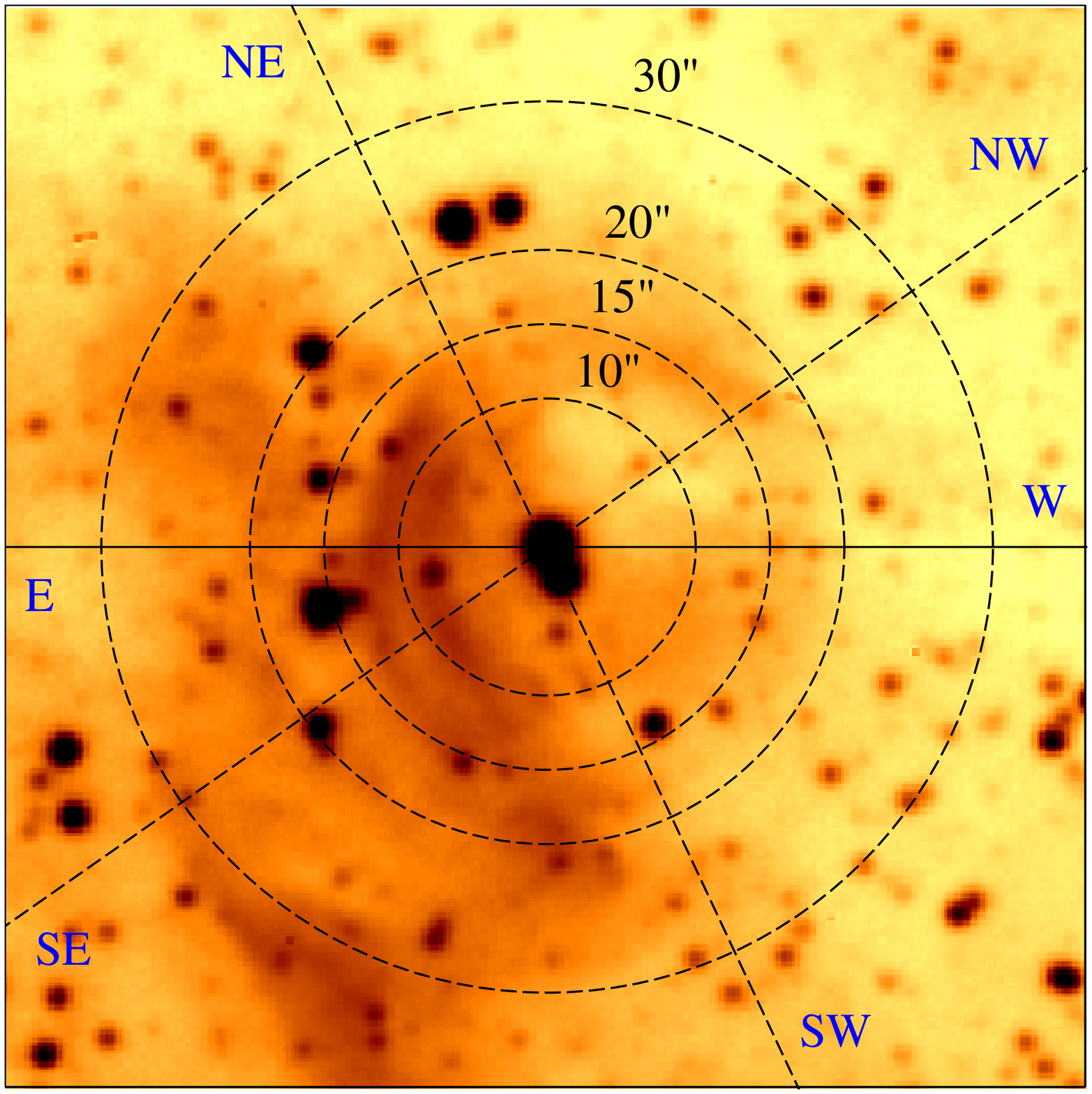}} {} \\
\end{minipage}
\caption{Upper panels: H$\alpha$ line intensity and velocity
profiles along the slit for two Gemini observations with
PA=125$\degr$ (left-hand panel) and PA=22$\degr$. SE--NW and
NE--SW directions of the slit are shown. The solid vertical line
corresponds to the position of BAT99\,3a, while the dashed
vertical lines (at $\pm$3 arcsec from the solid one) mark the
area, where the radial velocity was not measured because of the
effect of BAT99\,3a and star\,2. Bottom left-hand panel: He\,{\sc
ii} $\lambda$4686 line intensity profile along the slit for the
second du Pont observation with PA=90$\degr$. E--W direction of
the slit is shown. The solid vertical line corresponds to the
position of BAT99\,3a. The solid horizontal line shows the
background level and the dotted horizontal ones show $\pm1\sigma$
of the background noise. All data are background subtracted.
Bottom right-hand panel: MCELS2 H$\alpha$ image of the circular
shell with the slit positions for the three above-mentioned
observations indicated. Dashed lines correspond to the Gemini
slits and the solid one to the du Pont slit. Concentric, dashed
circles are over-plotted on the image to make its comparison with
the H$\alpha$ and He\,{\sc ii} $\lambda$4686 line intensities and
H$\alpha$ velocity profiles more convenient.} \label{fig:pv}
\end{figure*}

To study the gas kinematics around BAT99\,3a and star\,2, we
calculated the distribution of the heliocentric radial velocity,
$V_{\rm hel}$, along the slit using both Gemini spectra and the
method and programs described in Zasov et al. (2000). To exclude
possible systematic shifts, the closest bright night sky line
[O\,{\sc i}] $\lambda$6363 \AA \, was used. To match the scale
along the slit with the seeing during the observations, the
spectra were rebinned from 0.073 to 0.4 arcsec pixel$^{-1}$.
Finally, only those velocity measurements were used which satisfy
the criteria S/N\,$>$\,3 and $\sigma_v <5 \, \kms$.

The upper panels in Fig.\,\ref{fig:pv} plot the H$\alpha$
intensity and $V_{\rm hel}$ distributions along the slit for both
Gemini spectra, obtained for PA=125$\degr$ (left-hand panel) and
PA=22$\degr$. In both cases, the H$\alpha$ emission appears
everywhere along the slit, but its intensity shows clear
correlation with the shell (see the bottom right-hand panel for
the MCELS2 H$\alpha$ image of the shell with the slit positions
for the Gemini observations shown by dashed lines). Particularly,
it has two peaks in the southeast direction (at $\approx$8 and 13
arcsec from BAT99\,3a), which correspond to the two arcs on the
eastern side of the shell. In the northwest direction, the
H$\alpha$ emission peaks at $\approx$18 arcsec, which is
consistent with the greater extent of the shell in this direction.
One can also see that $V_{\rm hel}$ is quite uniform within the
shell and in the southeast and southwest directions, and does not
show a distinct correlation with the H$\alpha$ intensity peaks. On
the other hand, there is a noticeable increase of $V_{\rm hel}$ in
the northwest direction, and to a lesser extent in the northeast
one, suggesting that the shell might have a blister-like
structure.

The bottom left-hand panel in Fig.\,\ref{fig:pv} plots intensity
of the He\,{\sc ii} $\lambda$4686 line along the slit (indicated
in the bottom right-hand panel by a solid line with PA=90$\degr$)
for the du Pont spectrum taken on 2012 December 13. One can see
that the He\,{\sc ii} emission shows a good correlation with the
shell. Like the H$\alpha$ emission, it is the brightest on the
eastern edge of the shell and extends to a larger angular distance
in the western direction. On the other hand, the He\,{\sc ii}
emission is concentrated only in the shell, which implies that the
He\,{\sc iii} region is trapped by the shell. It is instructive,
therefore, to derive the effective temperature of the WR star by
applying the Zanstra method (Zanstra 1927). With $I$(He\,{\sc ii}
$\lambda$4686)/$I$(H$\beta$)$\approx$0.07 (see
Table\,\ref{tab:int3}) and using equation\,(1) in Kaler \& Jacoby
(1989), one finds $T_{\rm eff}$$\approx$95 kK, which agrees fairly
well with the temperature derived in Section\,\ref{sec:mod} (see
Table\,\ref{tab:model}).

\subsection{Origin of the shell}
\label{sec:neb-ori}

In Section\,\ref{sec:phys}, we found that the chemical composition
and electron number density of the shell are similar to those
measured for \hii regions and interstellar shells around WR stars,
which implies that the shell originated because of interaction
between the WR wind and the local interstellar
medium\footnote{Note that the total mechanical luminosity of the O
component of BAT99\,3a and star\,2 constitutes about 10 per cent
of the mechanical luminosity of the WN3 star.}. This implication
can be further supported by using diagnostic diagrams, which allow
to classify emission-line nebulae on the basis of line ratios (see
e.g. Kniazev, Pustilnik \& Zucker 2008a; Frew \& Parker 2010, and
references therein). With ratios
$\log$($I$(H$\alpha$)/$I$([N\,{\sc ii}]
$\lambda\lambda$6548,6584))=1.21, $\log(I$(H$\alpha$)/$I$([S\,{\sc
ii}] $\lambda\lambda$6717,6731))=0.84, and $\log(I$([O\,{\sc iii}]
$\lambda$5007)/$I$(H$\beta$))=0.86, the shell is located in the
region occupied by \hii regions and interstellar shells around
Magellanic Cloud WR stars (see fig. 4 in Frew \& Parker 2010). We
therefore conclude that the circular shell is a wind-blown
structure photoionized by its central stars.

We speculate that the two-arc morphology of the shell could be
caused by a sharp density gradient along the line-of-sight,
resulting in blow-up of the shell in the transverse direction as
it entered in the region of lower density. This, along with the
brightness asymmetry of the shell and the displacement of its
central stars, provides support to the possibility that the shell
is associated with a blister-like, wind-driven bubble.

\section{Discussion and conclusion}
\label{sec:rel}

There is mounting evidence that massive stars form in a clustered
way (Gvaramadze et al. 2012b and references therein).
Subsequently, some of them found themselves in the field because
of dynamical few-body encounters (Poveda et al. 1967; Gies \&
Bolton 1986), binary-supernova explosions (Blaauw 1961; Stone
1991) and cluster dissolution caused by gas expulsion (Tutukov
1978; Kroupa, Aarseth \& Hurley 2001). The discovery of two
massive and therefore rare stars, one of which in its turn is a
massive binary system, at $\approx$0.5 pc in projection from each
other, gives rise to the question of their physical relationship.

A chance projection along the same line-of-sight of physically
unrelated massive field stars can be rejected to a considerable
degree of reliability. Although such projections might be common
in the vicinity of massive star clusters, they are highly
improbable in the field (cf. Gvaramadze \& Menten 2012). One can
envisage two possible explanations of the close proximity of
BAT99\,3a and star\,2.

Firstly, these two stars might be members (at least in the recent
past) of a {\it runaway} triple (or higher multiplicity)
hierarchical system, which became unstable and dissolved into a
binary system (now BAT99\,3a) and a single unbound star (star\,2)
in the course of evolution of its most massive component(s) (e.g.
Gvaramadze \& Menten 2012). If this is the case, then the
brightness asymmetry of the shell and the offset of the stars
towards its brightest portion could be because of the eastward
motion of the system. For a typical space velocity of runaway
stars of several tens of $\kms$ and assuming that the system was
ejected $\sim$6 Myr ago (i.e. soon after its formation in the
parent cluster), one finds that the cluster should be at
$\sim$100$-$200 pc away. Proceeding from this, we searched for
known star clusters to the west of the shell and found two open
clusters, [HS66]\,49 and IC\,2111, at $\approx$ 41 and 92 pc in
projection, respectively (cf. Gvaramadze et al. 2012c). The age of
the former cluster is unknown, while that of the latter one is
6$\pm$1 Myr (Wolf et al. 2007), i.e. consistent with the age of
the WN3 star of $\approx$6 Myr (see Section\,\ref{sec:mod}). The
runaway scenario, however, implies that the circular shell should
be of circumstellar origin (cf. Lozinskaya 1992; Danforth \& Chu
2001; Gvaramadze et al. 2009), which is inconsistent with the
normal chemical composition of the shell. It is likely therefore
that the lopsided appearance of the shell and the displacement of
its central stars are simply because of the eastward density
gradient of the ambient medium (cf. Sections\,\ref{sec:nebula} and
\ref{sec:neb}).

Secondly, BAT99\,3a and star\,2 could be members of a yet
unrecognized star cluster. Observational and theoretical arguments
suggest that star clusters are formed in compact configurations
with the characteristic radius of $\leq$1 pc, which is independent
of cluster mass (Kroupa \& Boily 2002). This makes them hardly
resolved in the Magellanic Clouds, where 1 pc corresponds to
$\approx$3$-$4 arcsec. Indeed, some of the brightest stars in
these galaxies are today known to be star clusters (e.g. Weigelt
\& Baier 1985; Walborn et al. 1995, 1999; Heydari-Malayeri et al.
2002). Moreover, the detection of young star clusters could
additionally be hampered because of heavy obscuration by dust in
their parent molecular clouds. Some members of these embedded
clusters, however, could be visible in optical wavelengths if they
were dynamically ejected out of the obscuring material (Gvaramadze
et al. 2010b, 2012b) and/or if they are located in a favourably
oriented blister-like H\,{\sc ii} region on the near side of the
natal cloud (cf. Gvaramadze et al. 2011c). It is possible that
just this situation takes place in the case of BAT99\,3a and
star\,2 (cf. Section\,\ref{sec:mod}), and that their cousins in
the parent cluster are hidden by obscuring material along our
line-of-sight.

We found some support to this possibility in the $JHK_{\rm s}$
survey of the Magellanic Clouds by Kato et al. (2007). As we
already noted in Section\,\ref{sec:nebula}, this survey resolved
BAT99\,3a into two stars separated by 1 arcsec from each other
(see Fig.\,\ref{fig:acq} and Table\,\ref{tab:phot}). Assuming that
star\,3 is a massive member of the same cluster as BAT99\,3a and
star\,2, one can derive the $K$-band extinction towards it using
the $J$ and $K_{\rm s}$ magnitudes from Table\,\ref{tab:phot} and
the relationship:
\begin{equation}
A_K =0.66[(J-K)-(J-K)_0] \, , \label{eqn:AK}
\end{equation}
where $K$=$K_{\rm s}$+$0.04$ mag (Carpenter 2001). For the
intrinsic $(J-K)_0$ colour of $-$0.21 mag (typical of O stars;
Martins \& Plez 2006), we found $A_K$=0.61 mag and the absolute
$K$-band magnitude $M_K=-3.39$ mag, which corresponds to an
O8.5\,V star (Martins \& Plez 2006). Then using $M_V$=$-$4.25 mag
(Martins \& Plez 2006) and $A_V =A_K /0.112$ (Rieke \& Lebofsky
1985), we found $A_V$=5.42 mag and the visual magnitude of star\,3
of $V$$\approx$19.7 mag. The latter two estimates imply that the
extinction towards the putative cluster is variable, which is
often the case in young star clusters (e.g. Sagar, Munari \& de
Boer 2001; Espinoza, Selman \& Melnick 2009), and that the
contribution of star\,3 to the (optical) spectrum of BAT99\,3a is
negligible (cf. Section\,\ref{sec:mod}).

To conclude, follow-up IR spectroscopy of star\,3 is highly
desirable to confirm its status as an O star and to prove the
existence of a star cluster possibly associated with BAT99\,3a and
star\,2. The latter in its turn would provide one more piece of
evidence that massive stars form in a clustered mode.

\section{Acknowledgements}

Some observations reported in this paper were obtained with the
Southern African Large Telescope (SALT). ANC gratefully
acknowledges support from the Chilean Centro de Astrof\'{i}sica
FONDAP No.15010003, the Chilean Centro de Excelencia en
Astrof\'{i}sica y Tecnolog\'{i}as Afines (CATA) BASAL PFB-06/2007,
the Comite Mixto ESO-Gobierno de Chile and GEMINI-CONICYT No.
32110005. AYK acknowledges support from the National Research
Foundation of South Africa. This work has made use of the
NASA/IPAC Infrared Science Archive, which is operated by the Jet
Propulsion Laboratory, California Institute of Technology, under
contract with the National Aeronautics and Space Administration,
the SIMBAD data base and the VizieR catalogue access tool, both
operated at CDS, Strasbourg, France.

\end{document}